\documentclass[sigconf]{acmart}
\usepackage{amsmath}
\usepackage{enumitem}
\usepackage{numprint}
\npthousandsep{,}

\usepackage[normalem]{ulem} 

\AtBeginDocument{%
  \providecommand\BibTeX{{%
    \normalfont B\kern-0.5em{\scshape i\kern-0.25em b}\kern-0.8em\TeX}}}

\setcopyright{acmcopyright}
\copyrightyear{2018}
\acmYear{2018}
\acmDOI{10.1145/1122445.1122456}

\acmConference[Woodstock '18]{Woodstock '18: ACM Symposium on Neural
  Gaze Detection}{June 03--05, 2018}{Woodstock, NY}
\acmBooktitle{Woodstock '18: ACM Symposium on Neural Gaze Detection,
  June 03--05, 2018, Woodstock, NY}
\acmPrice{15.00}
\acmISBN{978-1-4503-XXXX-X/18/06}

\begin{document}

\title{Rethinking Lifelong Sequential Recommendation with Incremental Multi-Interest Attention}

\author{Yongji Wu$^{1*}$, Lu Yin$^{3}$, Defu Lian$^{2}$, Mingyang Yin$^{3}$,\\ Neil Zhenqiang Gong$^{1}$, Jingren Zhou$^{3}$, Hongxia Yang$^{3}$}

\affiliation{
\institution{$^{1}$Duke University, $^{2}$University of Science and Technology of China, $^{3}$Alibaba Group}
\country{}}
\affiliation{
\institution{$^1$\{yongji.wu769, neil.gong\}@duke.edu, $^2$liandefu@ustc.edu.cn, \\$^3$\{theodore.yl, hengyang.ymy, jingren.zhou, yang.yhx\}@alibaba-inc.com}
\country{}}

\renewcommand{\shortauthors}{Y. Wu, et al.}
\renewcommand{\authors}{Yongji Wu, Lu Yin, Defu Lian, Mingyang Yin, Neil Zhenqiang Gong, Jingren Zhou, Hongxia Yang}



\begin{abstract}
Sequential recommendation plays an increasingly important role in many e-commerce services such as display advertisement and online shopping. With the rapid development of these services in the last two decades, users have accumulated a massive amount of behavior data. Richer sequential behavior data has been proven to be of great value for sequential recommendation. However, traditional sequential models fail to handle users' lifelong sequences, as their linear computational and storage cost prohibits them from performing online inference. Recently, lifelong sequential modeling methods that borrow the idea of memory networks from NLP are proposed to address this issue. However, the RNN-based memory networks built upon intrinsically suffer from the inability to capture long-term dependencies and may instead be overwhelmed by the noise on extremely long behavior sequences. In addition, as the user's behavior sequence gets longer, more interests would be demonstrated in it. It is therefore crucial to model and capture the diverse interests of users. In order to tackle these issues, we propose a novel \textbf{\underline{l}}ifelong \textbf{\underline{i}}ncremental \textbf{\underline{m}}ulti-interest self \textbf{\underline{a}}ttention based sequential recommendation model, namely LimaRec. Our proposed method benefits from the carefully designed self-attention to identify relevant information from users' behavior sequences with different interests. It is still able to incrementally update users' representations for online inference, similarly to memory network based approaches. We extensively evaluate our method on four real-world datasets and demonstrate its superior performances compared to the state-of-the-art baselines.
\end{abstract}



\maketitle

{
\renewcommand{\thefootnote}{\fnsymbol{footnote}}
\footnotetext[1]{This work was conducted while he was a research intern at Alibaba Group.}
}

\section{Introduction}
With the rapid development of e-commerce businesses such as online shopping, display advertisement, and video streaming services, a huge amount of behavior data has been produced by users. This provides us with great opportunities to utilize the increasingly richer sequential behavior data, and learn better user representations for a variety of tasks such as CTR prediction and recommendation~\cite{ren2019lifelong,li2019multi,zhou2018deep}. The value of rich behavior history (long behavior sequences) has been demonstrated in~\cite{ren2019lifelong} for user modeling. We focus on the sequential recommendation task~\cite{kang2018self,li2019multi,cen2020controllable}, which plays a vital role in various online services, including the aforementioned ones.

However, modeling long sequences in sequential recommendation is not a trivial task, and it poses a series of great challenges for us. The traditional sequential recommendation models need to store the whole sequence and perform inference over it~\cite{covington2016deep,zhou2018deep,zhou2018deep,xiao2020deep}. These methods suffer from immense computational and storage cost as it scales linearly with the sequence length. The strict storage and latency constraints have therefore hampered us from employing long sequences for real-time inference in online systems~\cite{pi2020search}. Furthermore, as the user behavior sequence gets longer, it is more likely to contain items belonging to different categories, revealing the multi-facet nature of users' interests~\cite{ren2019lifelong, pi2020search}. The diverse interests of users also contribute to the complex temporal dynamics in long behavior sequences, where various temporal dependency patterns may exhibit~\cite{ren2019lifelong}. If we fail to identify the different interests manifested in the user's behavior history, we would instead be overwhelmed by noise brought by irrelevant behaviors using long sequences. 

Recently, a few pioneering studies have been conducted to address this issue under the click-through rate (CTR) prediction task. \cite{ren2019lifelong} first formulates the problem of lifelong sequential modeling. It borrows the idea of memory networks from the NLP domain, where a personalized behavior memory is maintained for each user to memorize his/her behavior sequences. The RNN-based hierarchical memory serves as the user representation and can be incrementally updated in constant time when the user clicks a new item. \citet{pi2019practice} propose another memory-based framework which adopts a different memory update strategy from NTM~\cite{graves2014neural}. Both of the two methods rely on the same set of RNN parameters to update the memory for each user. Therefore, they fail to adapt to each individual user's personalized behavioral transition patterns, as different users' behaviors have different temporal dynamics. Besides, RNN-based models pass sequential information through hidden states, they tend to forget about the past~\cite{bengio1994learning}. Though architectures like LSTM~\cite{bengio1994learning}, GRU~\cite{cho2014learning} and NTM~\cite{graves2014neural} have been proposed to alleviate this problem, they still suffer from the $O(l)$ path length between positions and fails to capture long-term dependencies if the sequence is too lengthy~\cite{kang2018self,vaswani2017attention}.

Search-based methods are introduced in~\cite{pi2020search,qin2020user} for CTR prediction. They consider that the whole behavior history contains much noise for the sequential models to directly make use of. Instead, they propose search strategies (a hand-crafted one in~\cite{pi2020search} and a model-based one in~\cite{qin2020user}) to retrieve only the relevant sub-sequence for the recommender to model. However, these methods utilize the target item to perform the search. Therefore, they cannot be applied to the sequential recommendation task, or the matching stage of recommendation systems, where target items are not available as model inputs.

The self-attention presented in Transformers~\cite{vaswani2017attention}  has seen incredibly successful in many sequential modeling tasks across a variety of fields, such as machine translation~\cite{xu2020layoutlm,devlin2018bert,dai2019transformer} and computer vision~\cite{wang2018non,parmar2018image,bello2019attention}. The self-attention mechanism uses the sequence to attend to itself. It adaptively assigns different weights to different positions of the sequence according to their relative importance, and then aggregates the sequence using the computed weights. Through this way, we can filter out the irrelevant behaviors in the user's historical sequence~\cite{kang2018self}. Compared to RNNs, the self-attention structure has a maximal path length of $O(1)$. This consequently facilitates self-attention in learning long-term dependencies. ~\cite{vaswani2017attention} Self-attention networks have also demonstrated superior performances in sequential recommendation~\cite{sun2019bert4rec,kang2018self,lian2020geography,zhang2018next}, which greatly outperform the RNN/CNN-based approaches.

Despite self-attention's triumph in sequential recommendation, and its intrinsic advantages in identifying relevant information from long sequences and capture long-term dependencies, we find that it is quite difficult to apply it to users' lifelong sequences in the online inference setting, since it bears the same aforementioned problem of linear computational and storage cost. Therefore it is not sufficient for online inference as we require the cost to be constant~\cite{ren2019lifelong,pi2019practice}. In the vanilla self-attention, when a new click action is generated by the user, we have to recompute the user representation using the entire sequence appended with the new action. Recently, linear self-attention methods have been introduced~\cite{katharopoulos2020transformers,choromanski2020rethinking}, where (linear) dot products of kernel feature maps are used as attention weights. We find that in our sequential recommendation scenario, the linear attention mechanism empowers us to perform incremental attention over the user's sequence. Through it, we can incrementally update the result of the self-attention operation as new click actions coming from users in constant time. This frees us from the need to store the user's whole historical sequence and enables us to perform real-time online inference, while still enjoys the benefits brought by the vanilla self-attention mechanism.

We observe that the attention mechanism can be considered as a soft-search operation where more relevant positions to a given query are assigned larger weight, hence they play more dominant roles in weighted aggregation. The attention mechanism therefore, naturally empowers us to capture users' multi-facet interests from their behavior sequences. As different behaviors of users are triggered by different underlying interests, we could utilize the attention mechanism to soft-search over the user's lifelong sequence and extract sub-sequences related to different interests of the user. Once again, to address the computational and storage inefficiency of the vanilla attention in online inference, we design a novel multi-interest extraction module under our incremental attention framework. Combined with the incremental self-attention blocks introduced in the previous paragraph, we present our novel \textbf{\underline{l}}ifelong \textbf{\underline{i}}ncremental \textbf{\underline{m}}ulti-interest self \textbf{\underline{a}}ttention based sequential recommendation model, namely LimaRec.

We summarize our contributions as follows:
\begin{itemize}[leftmargin=*]
    \item We propose a novel incremental self-attention based method for lifelong sequential recommendation, which goes beyond the limitations of RNN-based memory networks, while also possesses their ability to incrementally update the user representation for online inference. 
    \item Under the same incremental attention framework, we further propose a novel multi-interest extraction module to soft-search the whole sequence for the sub-sequences related to various interests of users.
    \item We conduct extensive experiments on four real-world datasets and obtain superior performance than state-of-the-art baselines. We also empirically investigate how does the performance of different methods vary on user sequences with increasing length.
\end{itemize}
\section{Related Work}
\subsection{Sequential Recommendation}
Early works on sequential recommendation usually utilize Markov chains (MCs) to capture the sequential patterns from users' sequential behaviors. FPMC~\cite{rendle2010factorizing} fuses MCs and matrix factorization to capture both long-term preferences and short-term transitions. Higher-order MCs are used to consider more previous items~\cite{he2016fusing,he2017translation}. With the advent of deep learning, deep sequential models like recurrent neural networks (RNNs) and convolutional neural networks (CNNs) have been extensively investigated in recent years and achieved great success~\cite{hidasi2015session,hidasi2018recurrent,tang2018personalized,chen2018sequential,li2017neural,liu2016context,yu2016dynamic}. 

Recently, self-attention networks~\cite{vaswani2017attention} have demonstrated superior performance in sequential recommendation due to full parallelism and the capacity to capture long-range dependencies. \cite{kang2018self} uses a binary cross-entropy loss based on inner product scores to train a self-attention network, while \citet{zhang2018next} propose to optimize a triplet margin loss based on Euclidean distance preference. The Cloze objective, which is proposed for training BERT~\cite{devlin2018bert}, is used to improve the performance sequential recommendation in~\cite{sun2019bert4rec}. However, these methods have to model the whole behavior sequence to perform inference and have a computational cost linear to the sequence length. Thus they can only be used to handle recent user behaviors in online inference.

\subsection{Lifelong Sequential Modeling}
The problem of lifelong sequential modeling is first introduced in~\cite{ren2019lifelong} for CTR prediction, where the authors propose a hierarchical periodic memory network (HPMN) to memorize users' behavior sequences. HPMN maintains hierarchical memories for each user to retain the information contained in his/her lifelong sequence. The memories are updated with different periods to capture the multi-scale sequential patterns of users. \cite{pi2019practice} is another method utilizing memory networks. Different from~\cite{ren2019lifelong}, which employs GRU to update the memory, \cite{pi2019practice} adopts the neural turning machine from~\cite{graves2014neural}. They further propose a memory utilization regularization strategy to reduce the variance of update weights across different memory slots, as well as a memory induction unit to capture high-order information. Compared to self-attention networks, RNNs have limited ability to capture long-term dependencies~\cite{vaswani2017attention}. This restrains the performance of the methods which rely on RNN-based memory networks.

On the other hand, \cite{pi2020search} and \cite{qin2020user} propose to address the issue from a different perspective. They argue that since it is computationally expensive to model the whole sequence, and there is much noise in the user's lifelong sequence, we should search for a limited number of the most relevant and appropriate user behaviors. The retrieved behaviors are then fed into a network to generate the final prediction. \cite{pi2020search} uses the category ID of items to find most relevant behaviors, while~\cite{qin2020user} learns a search strategy with reinforcement learning. However, both of the two methods are designed for CTR prediction, they rely on the information of the target item to perform the search. Since in sequential recommendation, the target item is not available in advance, these methods fail to adapt to our setting.

\subsection{Modeling Users' Diverse Interests}
The limit of using a single fixed-length representation to express users' multi-facet interests has been pointed out in~\cite{zhou2018deep}, where they introduced a local activation unit to adaptively learn the representations of user interests from behavior history, with respect to a given ad. DIEN~\cite{zhou2019deep} proposed to capture the interest evolving process of users with an interest extractor layer and an interest evolving layer, which use GRU as building blocks. A behavior refine layer is introduced in~\cite{xiao2020deep} to capture better user historical item representations for multi-interest extraction. The aforementioned methods are all devised for the CTR prediction task.

Distinguished from the multi-interest models for CTR prediction, \cite{li2019multi} proposes to model users' diverse interests in the sequential recommendation setting. They design a multi-interest extraction layer based on dynamic capsule routing from~\cite{sabour2017dynamic}. A different routing logic is used in~\cite{cen2020controllable} to extract interests, while another extraction method based on attention mechanism is also introduced. However, 
none of these methods is applicable in lifelong sequential recommendation, as they have a computational cost that scales linearly with the sequence length, while online inference requires $O(1)$ costs.
\newcommand\tp[2][-4]{{#2}^{\mkern#1mu\top}}

\section{Methodology}
\subsection{Preliminaries}

\begin{table}
    \caption{Important notations.}
    \label{tab:notations}
    \centering
    \resizebox{\columnwidth}{!}{
    \begin{tabular}{l p{0.8\linewidth}}
        \toprule
        Notation & Description\\
        \midrule
        $v_{l}$ & The $l$-th item clicked in the user's historical sequence. \\
        $\mathbf{e}_{l}$ & The embedding vector of the $l$-th item of the sequence. \\
        $\mathbf{p}_{l}$ & The positional embedding of the $l$-th position. \\
        $\mathbf{H}^{(i)}$ & The stacked latent representation of the sequence at each position after the $i$-th self-attention block. \\
        $\mathbf{h}^{(i)}_l$ & The latent representation up to the $l$-th position (i.e., the $l$-th row of $\mathbf{H}^{(i)}_l$). \\
        $\mathbf{S}^{(i)}$ & The stacked output of the $i$-th self-attention sub-layer (just before FFN). \\
        $\mathbf{s}^{(i)}_{l}$ & The $l$-th row of $\mathbf{S}^{i}$. \\
        $\mathbf{W}^{(i)}_{Q}, \mathbf{W}^{(i)}_{K}, \mathbf{W}^{(i)}_{V}$ & The key/query/value projection matrices of the $i$-th self-attention sub-layer. \\
        $\mathbf{r}^{(i)}_l, \mathbf{r}^{(i)}_l$ & The maintained hidden states for incremental attention of the $i$-th self-attention block at $l$-th step. \\
        $\widetilde{\mathbf{W}}_{K}, \widetilde{\mathbf{W}}_{V}$ & The key/value projection matrices of the multi-interest extraction module. \\        
        $\widetilde{\mathbf{r}}_l, \widetilde{\mathbf{z}}_l$ & The maintained hidden states for the multi-interest extraction module at the $l$-th step. \\
        $\mu_k$ & The trainable vector to encode the incentive behind the $k$-th interest. \\
        $\phi^{(k)}_l$ & The disentangled representation of the sequence under the $k$-th interest at the $l$-th step. \\
        \bottomrule
    \end{tabular}
    }
    \vspace{-1em}
\end{table}

\subsubsection{Problem Formulation}
For each user $u$, we have an ordered historical sequence of items $[v^{(u)}_1, v^{(u)}_2, \cdots, v^{(u)}_{L^{(u)}}]$ that $u$ has already interacted with (e.g., clicked), where $L^{(u)}$ is the current length of $u$'s historical sequence and $v^{(u)}_l$ is the index of the $l$-th item clicked by $u$. For simplicity, we omit the superscript $(u)$ from now. In the lifelong sequential recommendation setting, we consider that the user's sequence is dynamically updated according to the user's latest actions in the online serving scenario. The user $u$ may have clicked a new item $v_{L+1}$, which is appended to his/her historical sequence $\mathbf{v}$. Now we have the updated sequence $[v_1, v_2, \cdots, v_L, v_{L+1}]$. We then use this sequence to predict the next item (or the next set of items) $u$ might click. We consider the matching (candidate generation) stage of industrial recommendation systems~\cite{covington2016deep,li2019multi,cen2020controllable}, where we compute a preference score of user $u$ for each item $v$ based on the inner product similarity between the latent representation of $u$'s updated historical sequence (which includes $v_{L+1}$) and the embedding vector of $v$. The preference scores are then sorted to retrieve the top-$k$ items as the candidate set for $u$. When the user $u$ generates more actions $v_{L+2}, v_{L+3}, \cdots$, they are appended to the user's historical sequence, and the same recommendation procedure is repeated for online inference with respect to the user's real-time evolving behaviors.

Since we are handling with lifelong user sequences which might contain thousands of or even more behaviors, it would be infeasible to directly model the whole expanding user sequence in real-time online inference, due to the tremendous computational and storage costs. In addition, with the increasing sequence length, it becomes more and more difficult to precisely capture the complex multi-interests of users and the intricate temporal dynamics of users' evolving behavior patterns, without overwhelmed by massive noise. Therefore, lifelong sequential recommendation is a non-trivial task facing these challenges. To tackle these issues, we require the recommender to have the following properties under the lifelong sequential recommendation setting:
\begin{itemize}[leftmargin=*]
    \item The recommender should maintain a latent representation for each user's history in a fixed size (constant with respect to the sequence length), instead of storing the whole historical sequence.
    \item The maintained representation should efficiently preserve the  behavior information contained in the user's lifelong sequence. It should also capture the intrinsic multiple interests of the user and the involved temporal dynamics of the user's behaviors.
    \item The recommender should be able to continuously (incrementally) update the user representation to adapt to the new click behaviors generated by the user under this online setting.
\end{itemize}
Note that our task of lifelong sequential recommendation is different from that in~\cite{pi2019practice,ren2019lifelong}, where they focus on the CTR prediction task, in which the target item is used as model input. However, the target item is not required in our setting.

\subsubsection{Sequential Recommendation with Self-Attention}
\label{sec:sasrec}
\citet{kang2018self} introduce the SASRec model for sequential recommendation, which borrows the powerful self-attention mechanism from~\cite{vaswani2017attention}. Self-attention allows SASRec to identify the relevant behaviors in a user's history for next item prediction, through aggregating the sequence using adaptive weights.

In SASRec, two self-attention blocks are stacked to encode the user sequence. The user's historical sequence $[v_1, \cdots, v_L]$ is first transformed into a sequence of embedding vectors $[\mathbf{h}^{(0)}_1, \cdots, \mathbf{h}^{(0)}_L]$, where $\mathbf{h}^{(0)}_l \in \mathbb{R}^d = \mathbf{e}_l + \mathbf{p}_l$ is the sum of the embedding vector $\mathbf{e}_l$ of the $l$-th item and a trainable positional embedding $\mathbf{p}_l$. We then stack the embedding vectors $h^{(0)}_l$ into a matrix $\mathbf{H}^{(0)} \in \mathbb{R}^{L \times d}$. Self-attention is performed on $\mathbf{H}^{(0)}$, using each element of the sequence as query to attend to the sequence itself. Concretely, we compute the following output:
\begin{align}
\mathbf{S}^{(1)} &=\operatorname{Attn}(\mathbf{H}^{(0)} \tp{\mathbf{W}^{(1)}_Q}, \mathbf{H}^{(0)} \tp{\mathbf{W}^{(1)}_K}, \mathbf{H}^{(0)} \tp{\mathbf{W}^{(1)}_V}) \\
&= \operatorname{softmax} \left(\frac{(\mathbf{H}^{(0)} \tp{\mathbf{W}^{(1)}_Q})(\mathbf{H}^{(0)} \tp{\mathbf{W}^{(1)}_K})^{\top}}{\sqrt{d}}\right)\mathbf{H}^{(0)} \tp{\mathbf{W}^{(1)}_V}
\label{eq:sasrec_attention}
\end{align}
where $\mathbf{W}^{(1)}_Q, \mathbf{W}^{(1)}_K, \mathbf{W}^{(1)}_V \in \mathbb{R}^{d \times d}$ are the trainable projection matrices.

After the self-attention sub-layer, the output $\mathbf{S}^{(1)}$ is fed to a position-wise Feed-Forward Network (FFN) sub-layer, which is applied to each position in the sequence independently:
\begin{align}\label{eq:update_h}
\mathbf{h}^{(1)}_l = \operatorname{ReLU} ( \mathbf{W}^{(1)}_{FFN,1} \mathbf{s}^{(1)}_l + \mathbf{b}^{(1)}_{FFN,1}) \mathbf{W}^{(1)}_{FFN,2} + \mathbf{b}^{(1)}_{FFN,2}
\end{align}
where $ \mathbf{W}^{(1)}_{FFN,1},\mathbf{W}^{(1)}_{FFN,2} \in \mathbb{R}^{d\times d}$ and $\mathbf{b}^{(1)}_{FFN,1},\mathbf{b}^{(1)}_{FFN,2}\in \mathbb{R}^{d}$ are the parameters of the FFN sub-layer.

Subsequently, a second self-attention block is applied using $\mathbf{H}^{(1)}$ as input. A self-attention sub-layer performs the same computation as Eq.~\ref{eq:sasrec_attention} on $\mathbf{H}^{(1)}$, but with a different set of projection weights $\mathbf{W}^{(2)}_Q, \mathbf{W}^{(2)}_K, \mathbf{W}^{(2)}_V$. The result $\mathbf{S}^{(2)}$ then goes through another FFN sub-layer to produce the final output $\mathbf{H}^{(2)}$, where $\mathbf{h}^{(2)}_l$ is the latent representation of the user's historical sequence up to the $l$-th position. Residual connection and layer normalization are also used in SASRec to stabilize the network training.

We use $\mathbf{h}^{(2)}_L$ as the up-to-date user representation (since it encodes the entire historical sequence $[v_1, \cdots, v_L]$), and compute its inner product with $\mathbf{e}_j$ as the preference score for any item $j$.

\subsection{Modeling the Lifelong Sequence with Incremental Attention}
\label{sec:self_attention}
Through using the self-attention mechanism to both capture the long-term dependencies in the user's historical sequence, and focus only on the informative (relevant) part to reduce noise, SASRec outperforms state-of-the-art MC/CNN/RNN-based methods. However, a fatal problem arises naturally when we try to apply SASRec to lifelong sequences. Assume the user has clicked a new item $v_{L+1}$, then we have to update the user representation by computing the latent representation of the sequence appended with $v_{L+1}$. For the first self-attention block, we need to use $\mathbf{h}^{(0)}_{L+1}$ as query to attend to every position of the sequence:
\begin{align}
    \mathbf{s}^{(1)}_{L+1} &= \operatorname{softmax} \left(\frac{ (\mathbf{W}^{(i)}_Q \mathbf{h}^{(0}_{L+1})^{\top} (\mathbf{H}^{(0)} \tp{\mathbf{W}^{(1)}_K})^{\top}}{\sqrt{d}}\right)\mathbf{H}^{(0)} \tp{\mathbf{W}^{(1)}_V}\\
    & = \frac{\sum_{l^{\prime}=1}^{L} \exp \left( (\mathbf{W}^{(i)}_Q \mathbf{h}^{(0)}_{L+1})^{\top} (\mathbf{W}^{(i)}_K \mathbf{h}^{(0)}_{l^{\prime}}) \right) \mathbf{W}^{(i)}_V \mathbf{h}^{(0)}_{l^{\prime}}} {\sum_{l^{\prime}=1}^{L} \exp \left( (\mathbf{W}^{(i)}_Q \mathbf{h}^{(0)}_{L+1})^{\top} (\mathbf{W}^{(i)}_K \mathbf{h}^{(0)}_{l^{\prime}}) \right) }
\end{align}
Since we use softmax to normalize the scores, we have to compute the inner products (after projection) between $\mathbf{h}^{(0)}_{L+1}$ and $\mathbf{h}^{(0)}_{i}$ for $i=1,\ldots,L+1$, take the exponents, and then normalize them so that they are all positive and sum up to 1. Therefore, we must store the whole sequence $[\mathbf{h}^{(0)}_1, \cdots, \mathbf{h}^{(0)}_{L + 1}]$ in order to compute these inner products and perform the weighted average over this sequence. The same applies to the second self-attention block. This leads to a computational and storage cost of $O(L)$, which is prohibitive in the online setting of lifelong sequential recommendation. Note that when the new action $v_{L+1}$ arrives, we do not have to recompute $\mathbf{h}^{(i)}_1, \dots, \mathbf{h}^{(i)}_{L}$ (where $i=1, 2$), since we prevent previous positions from attending to subsequent positions. 

Recently, linear self-attention methods were introduced in to reduce the computational complexity of the self-attention operation from quadratic to linear. In these methods, the exponent similarity score (softmax) is replaced with kernels, the inner product of kernel feature maps $\varphi(\cdot)$ is used as the similarity score. Given queries, keys and values $\mathbf{Q}, \mathbf{K}, \mathbf{V} \in \mathbb{R}^{N \times d}$. The $i$-th output $\mathbf{V}_i$ is then computed as follows:
\begin{equation*}
    \mathbf{V}_i = \frac{\sum_{j=1}^N {\varphi(\mathbf{Q}_i)}^{\top} \varphi(\mathbf{K}_j) \mathbf{V}_j}{\sum_{j=1}^N {\varphi(\mathbf{Q}_i)}^{\top} \varphi(\mathbf{K}_j)}
\end{equation*}
According to the associativity of matrix multiplication, we can write it in the following form:
\begin{equation*}
    \mathbf{V}_i = \frac{{\varphi(\mathbf{Q}_i)}^{\top} \sum_{j=1}^N \varphi(\mathbf{K}_j) \mathbf{V}^{\top}_j}{{\varphi(\mathbf{Q}_i)}^{\top} \sum_{j=1}^N  \varphi(\mathbf{K}_j)}
\end{equation*}
Thus, we just need to compute the summations in the numerator and denominator once, and they can be repeatedly used for all quires.

\citet{choromanski2020rethinking} proposes the random feature map, $\varphi(\mathbf{x})=\frac{1}{\sqrt{m}} \exp (-\frac{\|\mathbf{x}\|^{2}}{2})(\exp (\omega_{1}^{\top} \mathbf{x}), \ldots, \exp (\omega_{m}^{\top} \mathbf{x}))$, to serve as an unbiased and low-variance approximation of the softmax self-attention. Here the vectors $\omega_{1}, \ldots, \omega_{m} \in \mathbb{R}^{d}$ are drawn i.i.d. from $\mathcal{N}(0, \mathbf{I}_{d})$.

In our setting, we restrict the $l$-th position to attend only to all the positions $l^{\prime}$ that $l^{\prime} \leq l$. Therefore, $\mathbf{s}^{(i)}_{l}$ can be computed as:
\begin{equation}
    \mathbf{s}^{(i)}_{l} = \frac{\varphi\left(\mathbf{W}^{(i)}_{Q} \mathbf{h}^{(i-1)}_{l}\right)^{\top} \sum_{l^{\prime}=1}^{l}
    \varphi\left(\mathbf{W}^{(i)}_{K}  \mathbf{h}^{(i-1)}_{l^{\prime}}\right) \left(\mathbf{W}^{(i)}_{V} \mathbf{h}^{(i-1)}_{l^{\prime}}\right)^{\top}
    }{\varphi\left(\mathbf{W}^{(i)}_{Q} \mathbf{h}^{(i-1)}_{l}\right) \sum_{l^{\prime}=1}^{l} \varphi\left(\mathbf{W}^{(i)}_{K}  \mathbf{h}^{(i-1)}_{l^{\prime}}\right) }
\end{equation}
We find that this restriction also empowers us the ability to perform attention over the whole sequence in an incremental manner. We only need to maintain two hidden states. When the sequence is updated with a new action $v_{L+1}$, we can compute the result of using $v_{L+1}$ to attend to the historical sequence from $v_{L+1}$ and the hidden states alone, without the need to store the entire sequence. Concretely, we maintain and incrementally update two hidden states $\mathbf{r}^{(i)}$ and $\mathbf{z}^{(i)}$ for the $i$-th self-attention block ($i=1, 2$) according to the following rules:
\begin{align}
    \mathbf{r}^{(i)}_l &=
        \begin{cases}\label{eq:update_r}
            0, \  & l=0\\
            \mathbf{r}^{(i)}_{l-1} + \varphi\left(\mathbf{W}^{(i)}_{K} \mathbf{h}^{(i-1)}_{l}\right) \left(\mathbf{W}^{(i)}_{V} \mathbf{h}^{(i-1)}_{l}\right)^{\top}, \  & l>0
        \end{cases} \\
    \mathbf{z}^{(i)}_l &=
        \begin{cases}\label{eq:update_z}
            0, \  & l=0\\
            \mathbf{z}^{(i)}_{l-1} + \varphi\left(\mathbf{W}^{(i)}_{K} \mathbf{h}^{(i-1)}_{l}\right), \  & l>0
        \end{cases}
\end{align}
Here $\mathbf{r}^{(i)}_{l}, \mathbf{z}^{(i)}_{l}$ are the states of $\mathbf{r}^{(i)}, \mathbf{z}^{(i)}$ at the $l$-th step (which considers the historical sequence up to the $l$-th action $v_{l}$). In this way, the output of the attention at the $l$-th position, which uses $v_{l}$ to query the sequence), can be computed as follows:
\begin{equation}
    \mathbf{s}^{(i)}_{l} = \frac{\varphi\left(\mathbf{W}^{(i)}_{Q} \mathbf{h}^{(i-1)}_{l} \right)^{\top} \mathbf{r}^{(i)}_{l} }{\varphi\left(\mathbf{W}^{(i)}_{Q} \mathbf{h}^{(i-1)}_{l}\right)^{\top} \mathbf{z}^{(i)}_{l} }
\label{eq:incremental_self_attn}
\end{equation}
As we can see, the computation of $\mathbf{s}^{(i)}_{l}$ depends only on the current output from previous layer $\mathbf{h}^{(i-1)}_{l}$ (note that $\mathbf{h}^{(0)}_{l}$ is the embedding of $v_l$) and the up-to-date hidden states $\mathbf{r}^{(i)}_{l}, \mathbf{z}^{(i)}_{l}$. The user sequence before the $l$-th position becomes irrelevant, and we can discard it. The output of the FFN sub-layer at each self-attention layer can be computed solely from $\mathbf{s}^{(i)}$. Therefore, for each user, the two hidden states of each self-attention block are all we need to store, and to dynamically update when a new action comes from the user. Since we use two self-attention blocks in our model, we need to maintain four hidden states in total for each user. Note that $\mathbf{r}^{(i)}\in\mathbb{R}^{d\times d}$ and $\mathbf{z}^{(i)}\in\mathbb{R}^{d}$. Although $\mathbf{r}^{(i)}$ is a $d\times d$ matrix, we have $L \gg d$ in our scenario, hence the $O(d^2)$ cost to store the hidden states is negligible compared to the $O(Ld)$ cost to store the whole sequence. Through maintain fixed-size states, we reduce the computational and storage complexity from $O(L)$ to $O(1)$ in online inference with newly arrived actions. 

Compared to the RNN-based memory networks~\cite{pi2019practice,ren2019lifelong}, which utilizes trainable gates to update the memory, our incremental attention based method aggregate all the positions of the historical sequence weighted by their "importance". The (unnormalized) score assigned to the $l^{\prime}$-th position is: $\varphi(\mathbf{W}^{(i)}_{Q} \mathbf{h}^{(i-1)}_{l})^{\top}\varphi(\mathbf{W}^{(i)}_{K} \mathbf{h}^{(i-1)}_{l^{\prime}})$. Self-attention possess the ability to filter out irrelevant part of the historical sequence (by assigning negligible weights)~\cite{kang2018self}, and our method enables it in an incremental fashion. This is particularly helpful for lifelong sequences, where the richer historical behavior data also leads to more noise. On the other hand, RNN-based methods may get overwhelmed by noise when modeling long sequences. Besides, they use the same set of RNN parameters to update the memory for each user. However, different users' behaviors have distinct transition patterns, using shared parameters fails to adapt to the personalized transition pattern of each individual user.

\subsection{Extracting Users' Multi-Facet Interests}
\label{sec:multi_interest_extract}
With the increasing length of the user's historical sequence, the more diverse the interests may contain in this sequence. Hence, if we directly model the whole sequence with a single representation, we cannot capture the multiple interests that the user has. Instead, we may got entangled with the various interests and fail to correctly identify the different driving forces behind the user's each action. Therefore, ideally, we should disentangle the user's whole sequence into multiple sub-sequences, each containing only the part related to a specific interest.

Coincidentally, we observe that the attention mechanism empowers us to complete this task. Computing the attention using the sequence itself as keys and values can be thought as performing a soft-search over the sequence with the given query, where the positions relevant to the query are assigned with larger weights, and they will dominate in the weighted average step. Therefore, the attention mechanism has the inherent capability to soft-extract the sub-sequences related to a query, and we use it as the building block of our proposed multi-interest extraction module.

We may characterize the various interests into a common collection that is shared among all the users, and each individual user may manifest a subset of it in his/her historical sequence. We assume the collection consists of $K$ potential interests of users, where $K$ is a tunable hyper-parameter. For each interest, we represent it with a trainable model parameter $\mu_k \in \mathbb{R}^{d}$, which encodes the underlying incentive behind this interest. $\mu_k$ is used to query the user sequence (after passing through the two linear self-attention blocks):
\begin{equation}\label{eq:disen_rep}
    \phi^{(k)}_{l} = \frac{\varphi\left(\mu_k \right)^{\top}  \sum_{l^{\prime}=1}^{l} \varphi\left(\widetilde{\mathbf{W}}_{K} \mathbf{h}^{(2)}_{l^{\prime}}\right) \left(\widetilde{\mathbf{W}}_{V} \mathbf{h}^{(2)}_{l^{\prime}}\right)^{\top}}{\varphi\left( \mu_k \right)^{\top} \sum_{l^{\prime}=1}^{l} \varphi\left(\widetilde{\mathbf{W}}_{K} \mathbf{h}^{(2)}_{l^{\prime}} \right) }
\end{equation}
Here $\phi^{(k)}_{l}$ represents the disentangled representation of the sequence at the $l$-th position (which encodes $[\mathbf{h}^{(2)}_{1}, \cdots, \mathbf{h}^{(2)}_{l}]$) under the $k$-th interest. $\widetilde{\mathbf{W}}_{K}, \widetilde{\mathbf{W}}_{V} \in \mathbb{R}^{d \times d}$ are the projection weights. Through linear attention, once again, we can incrementally update $\phi^{(k)}$, like what we do for the self-attention blocks in Section~\ref{sec:self_attention}. Since we use the same matrices to project keys and values for each interest, we only need to maintain two hidden states, which are shared among all the $K$ interests. The two states $\widetilde{\mathbf{r}}$  and $\widetilde{\mathbf{z}}$ are updated in the following manner:
\begin{align}
    \widetilde{\mathbf{r}}_l &=
        \begin{cases}
            0, \  & l=0\\
            \widetilde{\mathbf{r}}_{l-1} + \varphi\left(\widetilde{\mathbf{W}}_{K}  \mathbf{h}^{(2)}_{l}\right) \left(\widetilde{\mathbf{W}}_{V} \mathbf{h}^{(2)}_{l}\right)^{\top}, \  & l>0
        \end{cases} \\
    \widetilde{\mathbf{z}}_l &=
        \begin{cases}
            0, \  & l=0\\
            \widetilde{\mathbf{z}}_{l-1} + \varphi\left(\widetilde{\mathbf{W}}_{K} \mathbf{h}^{(2)}_{l}\right), \  & l>0
        \end{cases}
\end{align}
Armed with the up-to-date $\widetilde{\mathbf{r}}$  and $\widetilde{\mathbf{z}}$, we are enabled to compute $\phi^{(k)}_{l}$ as $(\varphi(\mu_k)^{\top}\widetilde{\mathbf{r}}_l) / (\varphi(\mu_k)^{\top}\widetilde{\mathbf{z}}_l)$ in $O(1)$ time with each incoming new behavior from the user, similar to Eq.~\ref{eq:incremental_self_attn}. The up-to-date set of vectors $\{\phi^{(k)}\}_{k=1}^{K}$ constitutes the final multi-interest representations of the current user sequence.

\subsection{Regularizing the Multi-Interest Representations}
The multi-interest extraction module we proposed in Section~\ref{sec:multi_interest_extract} conducts a soft search over the whole user sequence to retrieve the "sub-sequence" relevant to each interest. We would like the final representations $\{\phi^{(k)}\}_{k=1}^{K}$ to be disparate so that they can indeed capture the diverse interests of users. Otherwise, they may contain redundant information and $\{\mu_k\}$ may collapse to encode overlapping interests. Therefore, we should enforce $\{\phi^{(k)}\}_{k=1}^{K}$ to distill distinct information from the user sequence. We reckon that the target item (next item) should only be inferred from a single $\phi^{(k)}$ (corresponding to the one with the largest inner product score), since each user behavior should only be triggered by a single specific interest. Therefore, we propose the following regularization loss:
\begin{equation}
    \mathcal{L}_{\text{reg}} = \frac{\exp (\phi^{(k)} \mathbf{e}_{t})}{\sum_{k^{\prime}=1}^{K} \exp(\phi^{(k^{\prime})}\mathbf{e}_{t}) },
\end{equation}
where $k=\arg\max_{1 \leq k^{\prime} \leq K} (\phi^{(k^{\prime})} \mathbf{e}_{t})$ and $\mathbf{e}_t$ is the embedding of the target item. For the sequential recommendation task, we apply the binary cross-entropy loss:
\begin{equation}
    \mathcal{L}_{\text{rec}} = -\log \sigma(\phi^{(k)}\mathbf{e}_{t} ) - \mathbb{E}_{j \sim P_n} \log \left( 1 - \sigma(\phi^{(k)} \mathbf{e}_{j}) \right)
\end{equation}
Here we still denote $k=\arg\max_{1 \leq k^{\prime} \leq K} (\phi^{(k^{\prime})}\mathbf{e}_{t})$, and $P_n$ is a negative sampling distribution (we simply use a uniform one as in~\cite{kang2018self}).

The final loss function for a user is then given by:
\begin{equation}\label{eq:loss}
    \mathcal{L} = \mathcal{L}_{\text{rec}} + \lambda \mathcal{L}_{\text{reg}},
\end{equation}
where $\lambda$ is a hyperparameter controlling the weight of the regularization loss.

\section{Experiments}
\begin{table}
    \caption{Dataset statistics.}
    \label{tab:datasets}
    \centering
    \begin{tabular}{lrrrr}
        \toprule
        Dataset & \#users & \#items & \#actions & avg. length \\
        \midrule
        ML-1M & \numprint{6040} & \numprint{3416} & 1M & 165.50 \\
        ML-25M & \numprint{162542} & \numprint{36728} & 25M & 153.40 \\
        XLong & \numprint{20000} & \numprint{748471} & 25M & 796.40 \\
        Industrial & \numprint{430311} & \numprint{7093352} & 161M & 374.73 \\
        \bottomrule
    \end{tabular}
\end{table}

In this section, we empirically analyze the effectiveness of our proposed method and present the experimental results. We conduct the experiments in order to answer the following research questions:
\begin{itemize}
    \item [\textbf{RQ1:}] Does modeling lifelong sequences really contribute to the recommendation performance? How does LimaRec compare with state-of-the-art baselines?
    \item [\textbf{RQ2:}] How does the performance of different sequential models vary as the length of users' historical sequences increases?
    \item [\textbf{RQ3:}] Is it essential to consider the multi-facet interests of users in lifelong sequences? How does the number of interests we model influence the recommendation performance?
\end{itemize}

\subsection{Datasets}
We employ the following four real-world datasets to conduct experiments:
\begin{itemize}[leftmargin=*]
    \item MovieLens~\cite{harper2015movielens}: It is a series of frequently used datasets of movie ratings for evaluating recommendation systems. We use the two versions: MovieLens 1M (\textbf{ML-1M}) and MovieLens 25M (\textbf{ML-25M}), which include 1 million and 25 million ratings, respectively.
    \item \textbf{XLong}~\cite{ren2019lifelong}: The first public dataset tailored for lifelong sequential modeling. This dataset is collected from Alimama's online advertising platform, and contains relatively longer behavior sequences for each user.
    \item \textbf{Industrial}: A dataset sampled from user click logs on a top e-commerce platform within a period of one year.
\end{itemize}
The statistics of the datasets are shown in Table~\ref{tab:datasets}

\subsection{Competitors}
We compare the proposed LimaRec with a variety of baselines. The first group includes the state-of-the-art lifelong sequential modeling models:
\begin{itemize}[leftmargin=*]
    \item \textbf{MIMN}~\cite{pi2019practice} utilizes a memory network based on NTM~\cite{graves2014neural} to memorize the user's historical sequence. It improves the traditional NTM with a memory utilization regularization strategy and a memory induction unit.
    \item \textbf{HPMN}~\cite{ren2019lifelong} enhances memory networks with a hierarchical architecture with multiple update periods to mine multi-scale patterns in users' behavior sequence.
\end{itemize}
Note that these two methods are all developed for the CTR prediction task, where the personalized memory representation is concatenated with the feature vector of the target item and some user-side features. The concatenated vector is then fed into a multi-layer perceptron to produce the estimation of the user response probability. Since in our setting the target item is not available as model input, instead, we generate the candidate set by computing the inner product scores between the user's representation and the items. Hence, we modify MIMN and HPMN accordingly.

The second group contains the sequential recommendation methods designed to capture users' multiple interests:
\begin{itemize}[leftmargin=*]
    \item \textbf{MIND}~\cite{li2019multi} borrows the idea of dynamic capsule routing from~\cite{sabour2017dynamic}. Different from~\cite{sabour2017dynamic},  it employs shared bi-linear mapping matrix and randomly initialized routing logits.
    \item \textbf{ComiRec}~\cite{cen2020controllable} introduces two versions of multi-interest extraction modules. The first one, \textbf{ComiRec-DR}, follows the original routing mechanism in~\cite{sabour2017dynamic}. The second one, \textbf{ComiRec-SA}, explores a self-attentive mechanism. The attention mechanism is different from the scaled dot-product attention in SASRec.
\end{itemize}

The last group is composed of:
\begin{itemize}[leftmargin=*]
    \item \textbf{GRU4Rec}~\cite{hidasi2015session} is the first work that introduces RNNs to model user sequential behaviours. It is originally designed for session-based recommendation.
    \item \textbf{SASRec}~\cite{kang2018self}, the vanilla self-attention networks described in Section~\ref{sec:sasrec}.
\end{itemize}

\begin{table*}
    \centering
    \caption{Overall recommendation performance on MovieLens and XLong. The suffix "-40" and "-1000" indicates whether the method is trained and evaluated with lifelong sequences or using only recent behaviors. The ones without suffix use lifelong sequences.}
    \label{tab:overall_rec_performance}
    \resizebox{0.98\textwidth}{!}{
    \begin{tabular}{lcccccccccccc}
        \toprule
        & \multicolumn{4}{c}{ML-1M} & \multicolumn{4}{c}{ML-25M} & \multicolumn{4}{c}{XLong} \\
        \cmidrule(lr){2-5}
        \cmidrule(lr){6-9}
        \cmidrule(lr){10-13}
        & HR@5 & NDCG@5 & HR@10 & NDCG@10 & HR@5 & NDCG@5 & HR@10 & NDCG@10 & HR@5 & NDCG@5 & HR@10 & NDCG@10 \\
        \midrule
        GRU4Rec-40 & 0.2646 & 0.1847 & 0.3805 & 0.2221 & 0.8206 & 0.6526 & 0.9047 & 0.6801 & 0.1328 & 0.0820 & 0.1603 & 0.0910 \\
        GRU4Rec-1000 & 0.2576 & 0.1775 & 0.3770 & 0.2158 & 0.8141 & 0.6484 & 0.8967 & 0.6754 & 0.1326 & 0.0822 & 0.1673 & 0.0935 \\
        \midrule
        ComiRec-DR-40 & 0.4192 & 0.2945 & 0.5760 & 0.3451 & 0.8964 & 0.7526 & 0.9296 & 0.7637 & 0.4005 & 0.2625 & 0.5916 & 0.3241 \\
        ComiRec-DR-1000 & 0.3997 & 0.2777 & 0.5531 & 0.3271 & 0.8342 & 0.6657 & 0.8960 & 0.6860 & 0.2195 & 0.1627 & 0.2842 & 0.1834 \\
        ComiRec-SA-40 & 0.3901 & 0.2651 & 0.5531 & 0.3175 & 0.8094 & 0.6239 & 0.8992 & 0.6534 & 0.2977 & 0.1963 & 0.4898 & 0.2578 \\
        ComiRec-SA-1000 & 0.4169 & 0.2874 & 0.5765 & 0.3389 & 0.8189 & 0.6404 & 0.9046 & 0.6685 & 0.2306 & 0.1555 & 0.3712 & 0.2004 \\
        MIND-40 & 0.5232 & 0.3624 & 0.6911 & 0.4169 & 0.8783 & 0.7261 & 0.9331 & 0.7440 & 0.2106 & 0.1317 & 0.2607 & 0.1481 \\ 
        MIND-1000 & 0.5331 & 0.3693 & 0.6995 & 0.4233 & 0.8710 & 0.7229 & 0.9325 & 0.7429 & 0.1844 & 0.1205 & 0.2350 & 0.1367 \\
        \midrule 
        MIMN & 0.2414 & 0.1594 & 0.3747 & 0.2023 & 0.7878 & 0.5997 & 0.8902 & 0.6332 & 0.1966 & 0.1334 & 0.2385 & 0.1470 \\
        HPMN & 0.4561 & 0.3168 & 0.6149 & 0.3682 & 0.9283 & 0.8270 & 0.9559 & 0.8359 & 0.3700 & 0.2599 & 0.4796 & 0.2944 \\
        \midrule
        SASRec-40 & 0.6631 & 0.5026 & 0.7772 & 0.5397 & 0.9265 & 0.7936 & 0.9710 & 0.8083 & 0.3337 & 0.2359 & 0.4657 & 0.2783 \\
        SASRec-1000 & 0.6954 & 0.5428 & 0.8043 & 0.5786 & 0.9325 & 0.7982 & 0.9755 & 0.8124 & 0.6805 & 0.5282 & 0.8075 & 0.5675 \\
        LimaRec & \textbf{0.8699} & \textbf{0.6774} & \textbf{0.9606} & \textbf{0.7072} & \textbf{0.9699} & \textbf{0.8717} & \textbf{0.9867} & \textbf{0.8772} & \textbf{0.8207} & \textbf{0.6682} & \textbf{0.8931} & \textbf{0.6918}\\    
        \bottomrule
    \end{tabular}
    }
\end{table*}

\begin{table}
    \centering
    \caption{Overall recommendation performance on the Industrial dataset.}
    \label{tab:overall_rec_performance_industry}
    \resizebox{0.98\columnwidth}{!}{
    \begin{tabular}{lcccc}
        \toprule
        & HR@5 & NDCG@5 & HR@10 & NDCG@10 \\
        \midrule
        GRU4Rec-40 & 0.1772 & 0.1241 & 0.2258 & 0.1399 \\
        GRU4Rec-1000 & 0.1744 & 0.1274 & 0.2247 & 0.1436 \\
        \midrule
        ComiRec-DR-40 & 0.4616 & 0.3361 & 0.6145 & 0.3856   \\
        ComiRec-DR-1000 & 0.4158 & 0.3073 & 0.5523 & 0.3514 \\
        ComiRec-SA-40 & 0.4502 & 0.3284 & 0.5985 & 0.3761 \\
        ComiRec-SA-1000 & 0.2906 & 0.1964 & 0.4263 & 0.2401 \\
        MIND-40 & 0.2220 & 0.1580 & 0.2766 & 0.1758 \\ 
        MIND-1000 & 0.2443 & 0.1835 & 0.3096 & 0.2047  \\
        \midrule 
        MIMN & 0.2665 & 0.2003 & 0.3476 & 0.2265 \\
        HPMN & 0.3578 & 0.2961 & 0.4262 & 0.3181 \\
        \midrule
        SASRec-40 & 0.4725 & 0.3644 & 0.5782 & 0.3986 \\
        SASRec-1000 & 0.7701 & 0.6349 & 0.8610 & 0.6644 \\
        LimaRec & \textbf{0.8000} & \textbf{0.6555} & \textbf{0.8911} & \textbf{0.6851} \\
        \bottomrule
    \end{tabular}
    }
    \vspace{-1.5em}
\end{table}

\subsection{Settings}
We set the maximum length of users' lifelong sequences to 1000, as in~\cite{ren2019lifelong}, on all four datasets. For the methods that are not specially designed for lifelong sequential modeling, we also evaluate them using only recent 40 behaviors as users' history. In other words, they are trained and evaluated on sub-sequences with length up to 40, and user sequences with more than 40 behaviors are divided into non-overlapping chunks of length 40. By comparing the performance of these methods in the two settings, we can illustrate the importance and difficulty of incorporating lifelong sequences. We note that except GRU4Rec, HPMN, MIMN and our proposed LimaRec, the models trained using lifelong sequences are not practical in the online inference setting, due to the huge computational cost.

We implement our method and all the baselines with PyTorch. We train all models using the Adam optimizer with a learning rate of 0.001 and a batch size of 128. All methods are trained for a maximum of 500 epochs and the embedding dimension is set to 32. For our model, we use a regularization coefficient $\lambda$ of 0.01 and a dropout ratio of 0.1. The number of interests $K$ on each dataset is tuned from $\{1, 2, \cdots, 9\}$. The hyperparameters of the baseline methods are also tuned to the best on each dataset.

Following~\cite{cen2020controllable,kang2018self}, we evaluate the performance with two widely used metrics of ranking evaluation: Hit Rate (HR) and NDCG. HR@$k$ counts the proportion of times that the ground-truth target item is among the top-$k$, while NDCG@$k$ assigns larger weights on higher positions. We report the two metrics at $k=5$ and $k=10$. The last item of each user's behavior sequence is used for evaluating, and the remaining behaviors are used for training. We randomly generate 100 negative samples, pair them with the ground-truth target item for the compared methods to rank, as in~\cite{kang2018self}.

Note that LimaRec, HPMN and the multi-interest baselines yield multiple representation vectors $\{\phi{(k)}\}_{k=1}^{K}$ for each user. We find the one who has the largest inner product score with the target item $t$: $\phi^{(k)}=\max_{1 \leq k^{\prime} \leq K} (\phi^{(k^{\prime})} \mathbf{e}_{t})$, and use the universal $\phi^{(k)}$ to compute the scores for all the left negative items for \textit{more efficiency}. The exact evaluation setting would be to, for each negative item, compute the inner product using all $K$ vectors independently, and use the maximum one as the score. That is, we should use $\max_{1 \leq k^{\prime} \leq K} (\phi^{(k^{\prime})} \mathbf{e}_{j})$ as the score for negative item $j$, instead of using $\phi^{(k)} \mathbf{e}_{j}$, where $k=\arg\max_{1 \leq k^{\prime} \leq K} (\phi^{(k^{\prime})} \mathbf{e}_{t})$.

\begin{figure}
	\centering
	\includegraphics[width=.96\linewidth]{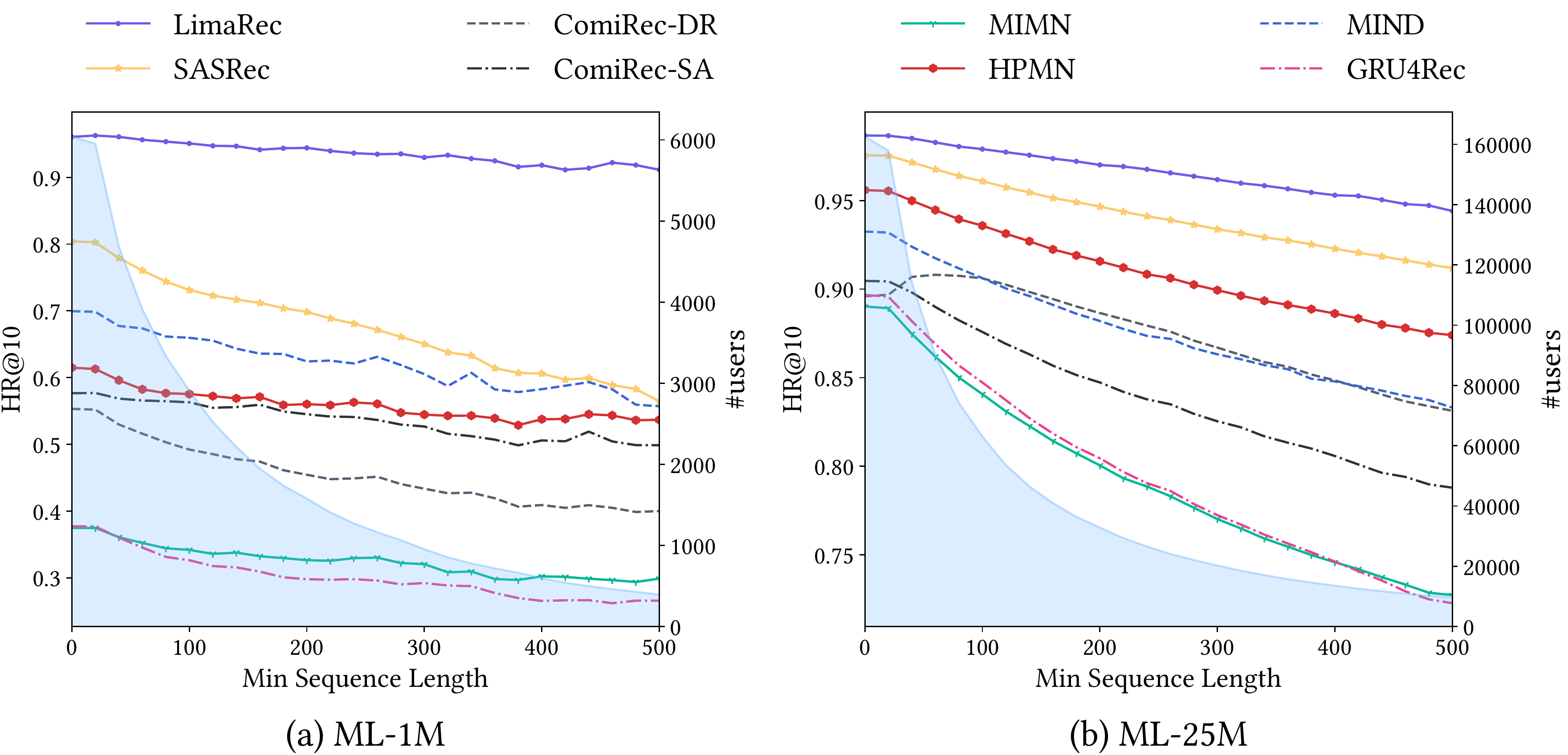}
	\caption{Recommendation performance on users with increasingly longer behavior sequences (MovieLens).}
	\label{fig:performance_wrt_length_movielens}
	\vspace{-2em}
\end{figure}

\begin{figure}
	\centering
	\includegraphics[width=.96\linewidth]{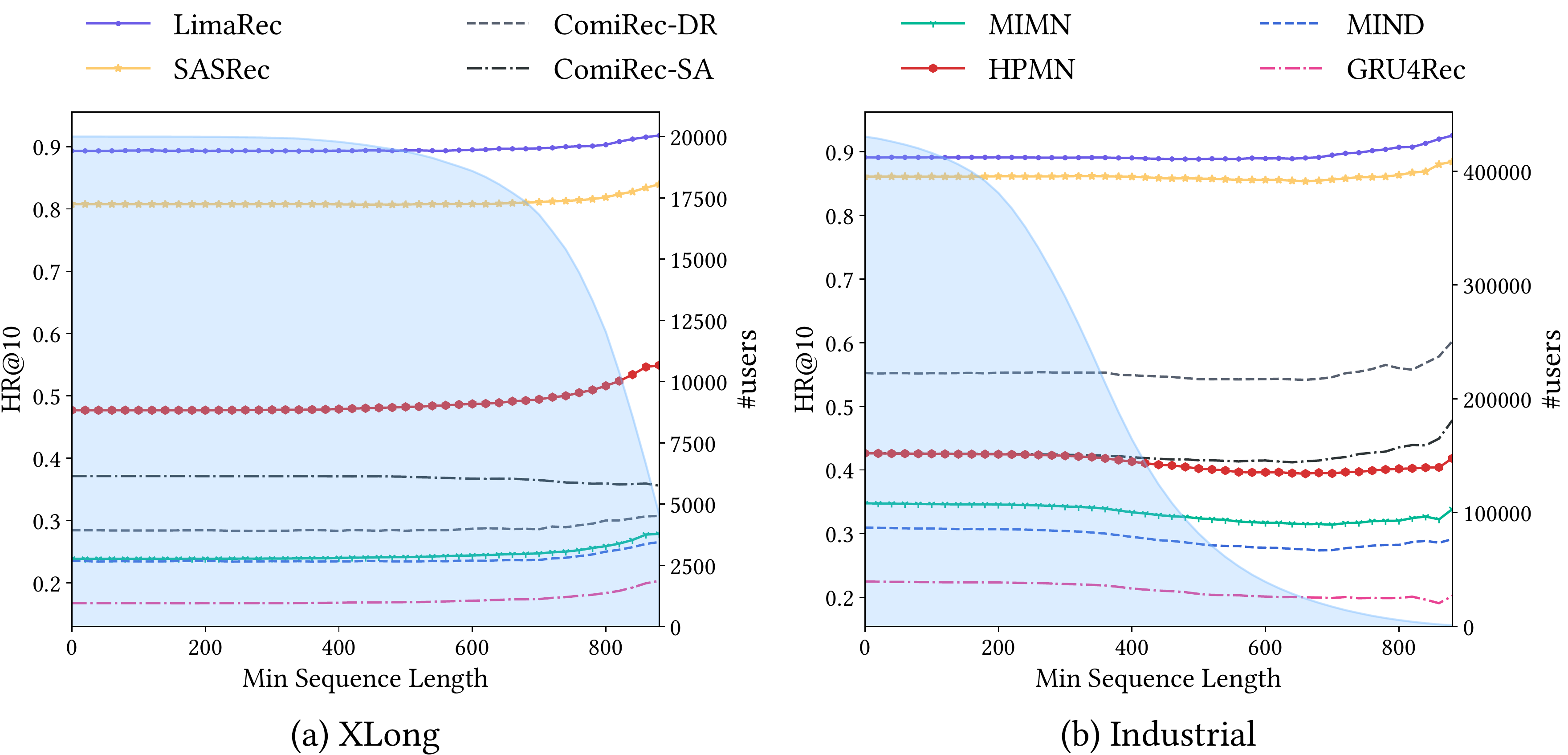}
	\caption{Recommendation performance on users with increasingly longer behavior sequences (XLong and Industrial).}
	\label{fig:performance_wrt_length_long}
	\vspace{-2em}
\end{figure}

\subsection{Overall Performance Comparison}
The results are summarized in Table~\ref{tab:overall_rec_performance} and Table~\ref{tab:overall_rec_performance_industry}. From the two tables, we have the following findings:
\begin{itemize}[leftmargin=*]
    \item \textit{Finding 1 - Incorporating lifelong sequences can boost the recommendation performance.} By comparing the performance of SASRec-40 with that of SASRec-1000, we find that training SASRec with lifelong sequences indeed improves the recommendation performance. Since the scaled dot-product self-attention can naturally identify relevant information in the user sequence with adaptive attention weights, SASRec can benefit from longer historical sequences without overwhelmed by noise. We also observe that HPMN outperforms the multi-interests models in most scenarios, this further illustrates the benefits brought by lifelong sequences.
    \item \textit{Finding 2 - It is not a simple task to modeling lifelong sequences.} We find that most multi-interests methods and GRU4Rec, which are not tailored for lifelong sequential modeling, however, decrease in performance in most cases when feeding lifelong sequences to them. These models fail to extract richer information, but instead get consumed by noise handling far longer sequences. ComiRec-SA performs better using lifelong sequences on two of the four datasets. This again demonstrates the merit of self-attention mechanism.
    \item \textit{Finding 3 - LimaRec consistently outperforms both the state-of-the-art lifelong sequential modeling baselines and the multi-interest models on all four datasets.} Compared with the lifelong sequential models and SASRec, this illustrates the superiority of our incremental attention mechanism over memory networks, and the effectiveness of our proposed multi-interest extraction module.
\end{itemize}

\subsection{Performance regarding Sequence Length}
\subsubsection{Settings} In this section, we evaluate how does the performance of different methods varies as the length of users' behavior sequences increases. Concretely, we extract the subset of users whose behavior sequences has length greater or equal a threshold $l$, we compute the performance of different methods only on this subset of users. We then observe how does the performance vary when we increase the threshold $l$. Note that when $l$ is increased, the subset contains fewer users. We present the results in Figure~\ref{fig:performance_wrt_length_movielens} and Figure~\ref{fig:performance_wrt_length_long}. All the compared methods are trained using lifelong sequences. The shaded region in light blue shows the cumulative density of the user sequence length's distribution.

\begin{figure}
	\centering
	\includegraphics[width=.96\linewidth]{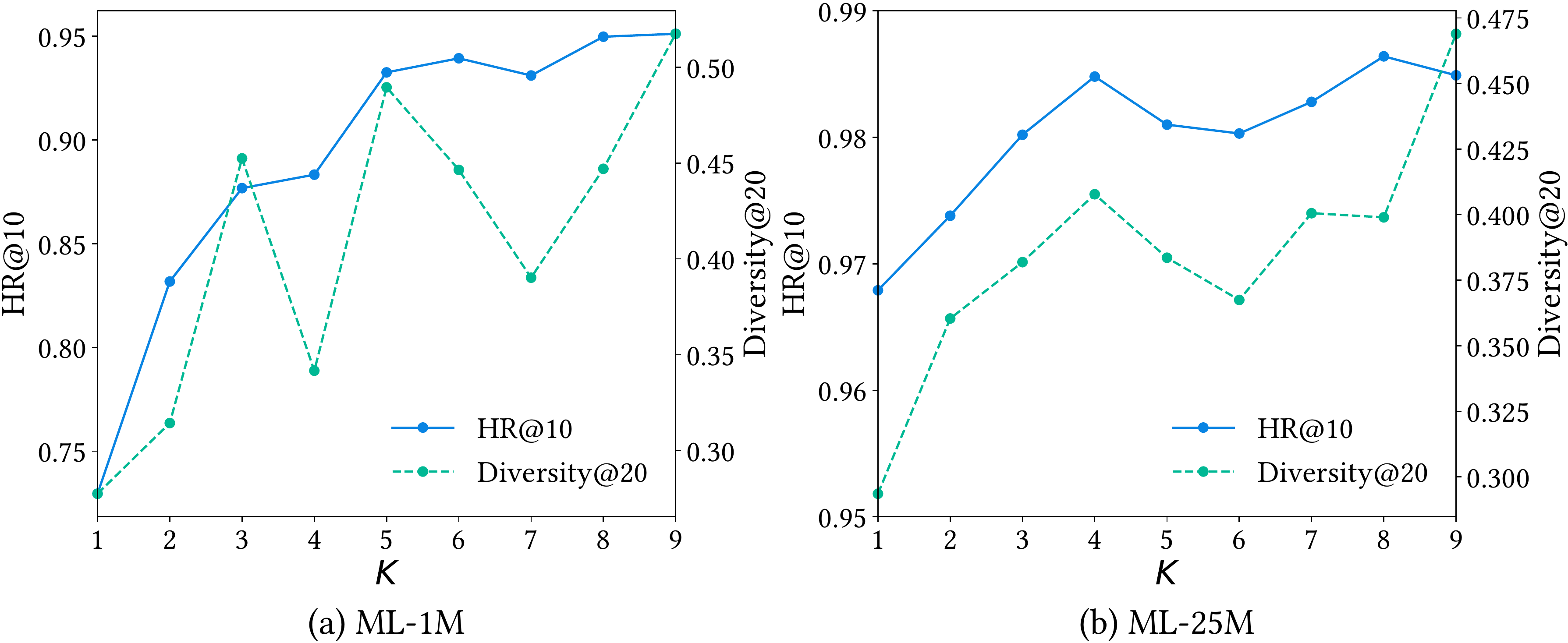}
	\caption{The impact of the number of interests $K$ on the hit rate and diversity.}
	\label{fig:interest_analysis}
	\vspace{-1em}
\end{figure}

\subsubsection{Findings} Surprisingly, we find that the recommendation performance varies differently on MovieLens and XLong/Industrial. From Figure~\ref{fig:performance_wrt_length_long}, we see that the performance would increase on users with longer sequences. This indicates that we can generate better recommendation to active users with rich behavior history, and further corroborates the importance of integrating lifelong sequences in sequential recommendation. On ML-1M and ML-25M, however, we observe that the performance drops on longer sequences. The MovieLens datasets are collected over the course of years (\textasciitilde3 yrs for ML-1M and \textasciitilde25 yrs for ML-25M). For a user with hundreds of actions, these actions may span a period of several years. On the other hand, the users with few actions may generate all of them within a short period, typically a day, and never log in again. The users with longer sequences are, in fact "sparser" than the ones with few behaviors. Hence, they contain much noise and would have fuzzy behavior transition patterns, and it is difficult to model them. Nevertheless, using lifelong sequences still contributes to the recommendation performance, as we see in Table~\ref{tab:overall_rec_performance}. Moreover, we see that on these two datasets, the performance of our proposed LimaRec decreases slower than the baseline methods when the sequence length increases, which indicates the superiority of LimaRec in modeling lifelong sequential behaviors.

\subsection{Analysis of the Multi-Interest Extraction Module}
\subsubsection{Settings} Here we empirically investigate the effectiveness of our proposed incremental attention based multi-interest extraction module. We vary the number of interests $K$ modeled in LimaRec from 1 to 9, train the model, and measure the hit rate. To better understand the nature of users' multi-facet interests, we also evaluate the recommendation diversity with respect to different $K$. We employ the diversity metric Diversity@$k$ in~\cite{cen2020controllable}, which computes the fraction of heterogeneous item-pairs (having different categories) in the top-$k$. The results are reported in Figure~\ref{fig:interest_analysis}.

\subsubsection{Findings} We observe that the hit rate using a single interest is significantly worse than utilizing multiple interests, as the lifelong sequences of users, by nature, contain diverse interests. When we set $K=1$, we are essentially using a single interest vector to query the latent representation of the sequence (after going through the two incremental self-attention blocks and projecting it to a new latent space). This would only amplify the noise contained in lifelong sequences. We obtain considerable performance boosts (in terms of hit rate) with increasing $K$. Hence it is quite essential to consider the multi-facet interests of users. Since it becomes easier for the model to recommend more diverse items with a larger $K$, we also find that, in general, the recommendation diversity tends to increase as $K$ increases. Though there are still some fluctuations when $K$ varies, it may be caused by the inconsistency between the $K$ we employed and the true number of interests in the user's historical sequence (which is unknown). One has to carefully tune $K$ to strike a balance between the traditional metrics like hit rate/NDCG, and the recommendation diversity.
\section{Conclusions}
In this work, we considered the problem of lifelong sequential recommendation, where the user representation has to be continuously updated in constant time to satisfy the strict latency and storage constraints for online systems, while still be able to preserve the behavior patterns in the historical sequence. Differing from the lifelong sequential modeling methods based on RNN memory networks, we propose a novel model built upon incremental self-attention. The self-attention mechanism enables us to identify and retrieve relevant information from the user's lifelong sequence, without succumbing to massive noise contained in it. We further propose a multi-interest extraction module to soft-search the lifelong sequence for the behaviors relevant to each interest. Extensive experiments have demonstrated the superiority of our method.
\bibliographystyle{ACM-Reference-Format}
\bibliography{reference}


\begin{thebibliography}{39}


\ifx \showCODEN    \undefined \def \showCODEN     #1{\unskip}     \fi
\ifx \showDOI      \undefined \def \showDOI       #1{#1}\fi
\ifx \showISBNx    \undefined \def \showISBNx     #1{\unskip}     \fi
\ifx \showISBNxiii \undefined \def \showISBNxiii  #1{\unskip}     \fi
\ifx \showISSN     \undefined \def \showISSN      #1{\unskip}     \fi
\ifx \showLCCN     \undefined \def \showLCCN      #1{\unskip}     \fi
\ifx \shownote     \undefined \def \shownote      #1{#1}          \fi
\ifx \showarticletitle \undefined \def \showarticletitle #1{#1}   \fi
\ifx \showURL      \undefined \def \showURL       {\relax}        \fi
\providecommand\bibfield[2]{#2}
\providecommand\bibinfo[2]{#2}
\providecommand\natexlab[1]{#1}
\providecommand\showeprint[2][]{arXiv:#2}

\bibitem[\protect\citeauthoryear{Bello, Zoph, Vaswani, Shlens, and Le}{Bello
  et~al\mbox{.}}{2019}]%
        {bello2019attention}
\bibfield{author}{\bibinfo{person}{Irwan Bello}, \bibinfo{person}{Barret Zoph},
  \bibinfo{person}{Ashish Vaswani}, \bibinfo{person}{Jonathon Shlens}, {and}
  \bibinfo{person}{Quoc~V Le}.} \bibinfo{year}{2019}\natexlab{}.
\newblock \showarticletitle{Attention augmented convolutional networks}. In
  \bibinfo{booktitle}{\emph{Proceedings of the IEEE/CVF International
  Conference on Computer Vision}}.
\newblock


\bibitem[\protect\citeauthoryear{Bengio, Simard, and Frasconi}{Bengio
  et~al\mbox{.}}{1994}]%
        {bengio1994learning}
\bibfield{author}{\bibinfo{person}{Yoshua Bengio}, \bibinfo{person}{Patrice
  Simard}, {and} \bibinfo{person}{Paolo Frasconi}.}
  \bibinfo{year}{1994}\natexlab{}.
\newblock \showarticletitle{Learning long-term dependencies with gradient
  descent is difficult}.
\newblock \bibinfo{journal}{\emph{IEEE transactions on neural networks}}
  \bibinfo{volume}{5}, \bibinfo{number}{2} (\bibinfo{year}{1994}).
\newblock


\bibitem[\protect\citeauthoryear{Cen, Zhang, Zou, Zhou, Yang, and Tang}{Cen
  et~al\mbox{.}}{2020}]%
        {cen2020controllable}
\bibfield{author}{\bibinfo{person}{Yukuo Cen}, \bibinfo{person}{Jianwei Zhang},
  \bibinfo{person}{Xu Zou}, \bibinfo{person}{Chang Zhou},
  \bibinfo{person}{Hongxia Yang}, {and} \bibinfo{person}{Jie Tang}.}
  \bibinfo{year}{2020}\natexlab{}.
\newblock \showarticletitle{Controllable Multi-Interest Framework for
  Recommendation}. In \bibinfo{booktitle}{\emph{Proceedings of the 26th ACM
  SIGKDD International Conference on Knowledge Discovery \& Data Mining}}.
\newblock


\bibitem[\protect\citeauthoryear{Chen, Xu, Zhang, Tang, Cao, Qin, and Zha}{Chen
  et~al\mbox{.}}{2018}]%
        {chen2018sequential}
\bibfield{author}{\bibinfo{person}{Xu Chen}, \bibinfo{person}{Hongteng Xu},
  \bibinfo{person}{Yongfeng Zhang}, \bibinfo{person}{Jiaxi Tang},
  \bibinfo{person}{Yixin Cao}, \bibinfo{person}{Zheng Qin}, {and}
  \bibinfo{person}{Hongyuan Zha}.} \bibinfo{year}{2018}\natexlab{}.
\newblock \showarticletitle{Sequential recommendation with user memory
  networks}. In \bibinfo{booktitle}{\emph{Proceedings of the eleventh ACM
  international conference on web search and data mining}}.
  \bibinfo{pages}{108--116}.
\newblock


\bibitem[\protect\citeauthoryear{Cho, Van~Merri{\"e}nboer, Gulcehre, Bahdanau,
  Bougares, Schwenk, and Bengio}{Cho et~al\mbox{.}}{2014}]%
        {cho2014learning}
\bibfield{author}{\bibinfo{person}{Kyunghyun Cho}, \bibinfo{person}{Bart
  Van~Merri{\"e}nboer}, \bibinfo{person}{Caglar Gulcehre},
  \bibinfo{person}{Dzmitry Bahdanau}, \bibinfo{person}{Fethi Bougares},
  \bibinfo{person}{Holger Schwenk}, {and} \bibinfo{person}{Yoshua Bengio}.}
  \bibinfo{year}{2014}\natexlab{}.
\newblock \showarticletitle{Learning phrase representations using RNN
  encoder-decoder for statistical machine translation}.
\newblock \bibinfo{journal}{\emph{arXiv preprint arXiv:1406.1078}}
  (\bibinfo{year}{2014}).
\newblock


\bibitem[\protect\citeauthoryear{Choromanski, Likhosherstov, Dohan, Song, Gane,
  Sarlos, Hawkins, Davis, Mohiuddin, Kaiser, et~al\mbox{.}}{Choromanski
  et~al\mbox{.}}{2020}]%
        {choromanski2020rethinking}
\bibfield{author}{\bibinfo{person}{Krzysztof Choromanski},
  \bibinfo{person}{Valerii Likhosherstov}, \bibinfo{person}{David Dohan},
  \bibinfo{person}{Xingyou Song}, \bibinfo{person}{Andreea Gane},
  \bibinfo{person}{Tamas Sarlos}, \bibinfo{person}{Peter Hawkins},
  \bibinfo{person}{Jared Davis}, \bibinfo{person}{Afroz Mohiuddin},
  \bibinfo{person}{Lukasz Kaiser}, {et~al\mbox{.}}}
  \bibinfo{year}{2020}\natexlab{}.
\newblock \showarticletitle{Rethinking attention with performers}.
\newblock \bibinfo{journal}{\emph{arXiv preprint arXiv:2009.14794}}
  (\bibinfo{year}{2020}).
\newblock


\bibitem[\protect\citeauthoryear{Covington, Adams, and Sargin}{Covington
  et~al\mbox{.}}{2016}]%
        {covington2016deep}
\bibfield{author}{\bibinfo{person}{Paul Covington}, \bibinfo{person}{Jay
  Adams}, {and} \bibinfo{person}{Emre Sargin}.}
  \bibinfo{year}{2016}\natexlab{}.
\newblock \showarticletitle{Deep neural networks for youtube recommendations}.
  In \bibinfo{booktitle}{\emph{Proceedings of the 10th ACM conference on
  recommender systems}}.
\newblock


\bibitem[\protect\citeauthoryear{Dai, Yang, Yang, Carbonell, Le, and
  Salakhutdinov}{Dai et~al\mbox{.}}{2019}]%
        {dai2019transformer}
\bibfield{author}{\bibinfo{person}{Zihang Dai}, \bibinfo{person}{Zhilin Yang},
  \bibinfo{person}{Yiming Yang}, \bibinfo{person}{Jaime~G Carbonell},
  \bibinfo{person}{Quoc Le}, {and} \bibinfo{person}{Ruslan Salakhutdinov}.}
  \bibinfo{year}{2019}\natexlab{}.
\newblock \showarticletitle{Transformer-XL: Attentive Language Models beyond a
  Fixed-Length Context}. In \bibinfo{booktitle}{\emph{Proceedings of the 57th
  Annual Meeting of the Association for Computational Linguistics}}.
\newblock


\bibitem[\protect\citeauthoryear{Devlin, Chang, Lee, and Toutanova}{Devlin
  et~al\mbox{.}}{2018}]%
        {devlin2018bert}
\bibfield{author}{\bibinfo{person}{Jacob Devlin}, \bibinfo{person}{Ming-Wei
  Chang}, \bibinfo{person}{Kenton Lee}, {and} \bibinfo{person}{Kristina
  Toutanova}.} \bibinfo{year}{2018}\natexlab{}.
\newblock \showarticletitle{Bert: Pre-training of deep bidirectional
  transformers for language understanding}.
\newblock \bibinfo{journal}{\emph{arXiv preprint arXiv:1810.04805}}
  (\bibinfo{year}{2018}).
\newblock


\bibitem[\protect\citeauthoryear{Graves, Wayne, and Danihelka}{Graves
  et~al\mbox{.}}{2014}]%
        {graves2014neural}
\bibfield{author}{\bibinfo{person}{Alex Graves}, \bibinfo{person}{Greg Wayne},
  {and} \bibinfo{person}{Ivo Danihelka}.} \bibinfo{year}{2014}\natexlab{}.
\newblock \showarticletitle{Neural turing machines}.
\newblock \bibinfo{journal}{\emph{arXiv preprint arXiv:1410.5401}}
  (\bibinfo{year}{2014}).
\newblock


\bibitem[\protect\citeauthoryear{Harper and Konstan}{Harper and
  Konstan}{2015}]%
        {harper2015movielens}
\bibfield{author}{\bibinfo{person}{F~Maxwell Harper} {and}
  \bibinfo{person}{Joseph~A Konstan}.} \bibinfo{year}{2015}\natexlab{}.
\newblock \showarticletitle{The movielens datasets: History and context}.
\newblock \bibinfo{journal}{\emph{Acm transactions on interactive intelligent
  systems (tiis)}} \bibinfo{volume}{5}, \bibinfo{number}{4}
  (\bibinfo{year}{2015}), \bibinfo{pages}{1--19}.
\newblock


\bibitem[\protect\citeauthoryear{He, Kang, and McAuley}{He
  et~al\mbox{.}}{2017}]%
        {he2017translation}
\bibfield{author}{\bibinfo{person}{Ruining He}, \bibinfo{person}{Wang-Cheng
  Kang}, {and} \bibinfo{person}{Julian McAuley}.}
  \bibinfo{year}{2017}\natexlab{}.
\newblock \showarticletitle{Translation-based recommendation}. In
  \bibinfo{booktitle}{\emph{Proceedings of the eleventh ACM conference on
  recommender systems}}. \bibinfo{pages}{161--169}.
\newblock


\bibitem[\protect\citeauthoryear{He and McAuley}{He and McAuley}{2016}]%
        {he2016fusing}
\bibfield{author}{\bibinfo{person}{Ruining He} {and} \bibinfo{person}{Julian
  McAuley}.} \bibinfo{year}{2016}\natexlab{}.
\newblock \showarticletitle{Fusing similarity models with markov chains for
  sparse sequential recommendation}. In \bibinfo{booktitle}{\emph{2016 IEEE
  16th International Conference on Data Mining (ICDM)}}. IEEE,
  \bibinfo{pages}{191--200}.
\newblock


\bibitem[\protect\citeauthoryear{Hidasi and Karatzoglou}{Hidasi and
  Karatzoglou}{2018}]%
        {hidasi2018recurrent}
\bibfield{author}{\bibinfo{person}{Bal{\'a}zs Hidasi} {and}
  \bibinfo{person}{Alexandros Karatzoglou}.} \bibinfo{year}{2018}\natexlab{}.
\newblock \showarticletitle{Recurrent neural networks with top-k gains for
  session-based recommendations}. In \bibinfo{booktitle}{\emph{Proceedings of
  the 27th ACM international conference on information and knowledge
  management}}. \bibinfo{pages}{843--852}.
\newblock


\bibitem[\protect\citeauthoryear{Hidasi, Karatzoglou, Baltrunas, and
  Tikk}{Hidasi et~al\mbox{.}}{2015}]%
        {hidasi2015session}
\bibfield{author}{\bibinfo{person}{Bal{\'a}zs Hidasi},
  \bibinfo{person}{Alexandros Karatzoglou}, \bibinfo{person}{Linas Baltrunas},
  {and} \bibinfo{person}{Domonkos Tikk}.} \bibinfo{year}{2015}\natexlab{}.
\newblock \showarticletitle{Session-based recommendations with recurrent neural
  networks}.
\newblock \bibinfo{journal}{\emph{arXiv preprint arXiv:1511.06939}}
  (\bibinfo{year}{2015}).
\newblock


\bibitem[\protect\citeauthoryear{Kang and McAuley}{Kang and McAuley}{2018}]%
        {kang2018self}
\bibfield{author}{\bibinfo{person}{Wang-Cheng Kang} {and}
  \bibinfo{person}{Julian McAuley}.} \bibinfo{year}{2018}\natexlab{}.
\newblock \showarticletitle{Self-attentive sequential recommendation}. In
  \bibinfo{booktitle}{\emph{2018 IEEE International Conference on Data Mining
  (ICDM)}}.
\newblock


\bibitem[\protect\citeauthoryear{Katharopoulos, Vyas, Pappas, and
  Fleuret}{Katharopoulos et~al\mbox{.}}{2020}]%
        {katharopoulos2020transformers}
\bibfield{author}{\bibinfo{person}{Angelos Katharopoulos},
  \bibinfo{person}{Apoorv Vyas}, \bibinfo{person}{Nikolaos Pappas}, {and}
  \bibinfo{person}{Fran{\c{c}}ois Fleuret}.} \bibinfo{year}{2020}\natexlab{}.
\newblock \showarticletitle{Transformers are rnns: Fast autoregressive
  transformers with linear attention}. In
  \bibinfo{booktitle}{\emph{International Conference on Machine Learning}}.
  PMLR, \bibinfo{pages}{5156--5165}.
\newblock


\bibitem[\protect\citeauthoryear{Li, Liu, Wu, Xu, Zhao, Huang, Kang, Chen, Li,
  and Lee}{Li et~al\mbox{.}}{2019}]%
        {li2019multi}
\bibfield{author}{\bibinfo{person}{Chao Li}, \bibinfo{person}{Zhiyuan Liu},
  \bibinfo{person}{Mengmeng Wu}, \bibinfo{person}{Yuchi Xu},
  \bibinfo{person}{Huan Zhao}, \bibinfo{person}{Pipei Huang},
  \bibinfo{person}{Guoliang Kang}, \bibinfo{person}{Qiwei Chen},
  \bibinfo{person}{Wei Li}, {and} \bibinfo{person}{Dik~Lun Lee}.}
  \bibinfo{year}{2019}\natexlab{}.
\newblock \showarticletitle{Multi-interest network with dynamic routing for
  recommendation at Tmall}. In \bibinfo{booktitle}{\emph{Proceedings of the
  28th ACM International Conference on Information and Knowledge Management}}.
\newblock


\bibitem[\protect\citeauthoryear{Li, Ren, Chen, Ren, Lian, and Ma}{Li
  et~al\mbox{.}}{2017}]%
        {li2017neural}
\bibfield{author}{\bibinfo{person}{Jing Li}, \bibinfo{person}{Pengjie Ren},
  \bibinfo{person}{Zhumin Chen}, \bibinfo{person}{Zhaochun Ren},
  \bibinfo{person}{Tao Lian}, {and} \bibinfo{person}{Jun Ma}.}
  \bibinfo{year}{2017}\natexlab{}.
\newblock \showarticletitle{Neural attentive session-based recommendation}. In
  \bibinfo{booktitle}{\emph{Proceedings of the 2017 ACM on Conference on
  Information and Knowledge Management}}. \bibinfo{pages}{1419--1428}.
\newblock


\bibitem[\protect\citeauthoryear{Lian, Wu, Ge, Xie, and Chen}{Lian
  et~al\mbox{.}}{2020}]%
        {lian2020geography}
\bibfield{author}{\bibinfo{person}{Defu Lian}, \bibinfo{person}{Yongji Wu},
  \bibinfo{person}{Yong Ge}, \bibinfo{person}{Xing Xie}, {and}
  \bibinfo{person}{Enhong Chen}.} \bibinfo{year}{2020}\natexlab{}.
\newblock \showarticletitle{Geography-aware sequential location
  recommendation}. In \bibinfo{booktitle}{\emph{Proceedings of the 26th ACM
  SIGKDD International Conference on Knowledge Discovery \& Data Mining}}.
\newblock


\bibitem[\protect\citeauthoryear{Likhosherstov, Choromanski, Davis, Song, and
  Weller}{Likhosherstov et~al\mbox{.}}{2020}]%
        {likhosherstov2020sub}
\bibfield{author}{\bibinfo{person}{Valerii Likhosherstov},
  \bibinfo{person}{Krzysztof Choromanski}, \bibinfo{person}{Jared Davis},
  \bibinfo{person}{Xingyou Song}, {and} \bibinfo{person}{Adrian Weller}.}
  \bibinfo{year}{2020}\natexlab{}.
\newblock \showarticletitle{Sub-Linear Memory: How to Make Performers SLiM}.
\newblock \bibinfo{journal}{\emph{arXiv preprint arXiv:2012.11346}}
  (\bibinfo{year}{2020}).
\newblock


\bibitem[\protect\citeauthoryear{Liu, Wu, Wang, Li, and Wang}{Liu
  et~al\mbox{.}}{2016}]%
        {liu2016context}
\bibfield{author}{\bibinfo{person}{Qiang Liu}, \bibinfo{person}{Shu Wu},
  \bibinfo{person}{Diyi Wang}, \bibinfo{person}{Zhaokang Li}, {and}
  \bibinfo{person}{Liang Wang}.} \bibinfo{year}{2016}\natexlab{}.
\newblock \showarticletitle{Context-aware sequential recommendation}. In
  \bibinfo{booktitle}{\emph{2016 IEEE 16th International Conference on Data
  Mining (ICDM)}}. IEEE, \bibinfo{pages}{1053--1058}.
\newblock


\bibitem[\protect\citeauthoryear{Parmar, Vaswani, Uszkoreit, Kaiser, Shazeer,
  Ku, and Tran}{Parmar et~al\mbox{.}}{2018}]%
        {parmar2018image}
\bibfield{author}{\bibinfo{person}{Niki Parmar}, \bibinfo{person}{Ashish
  Vaswani}, \bibinfo{person}{Jakob Uszkoreit}, \bibinfo{person}{Lukasz Kaiser},
  \bibinfo{person}{Noam Shazeer}, \bibinfo{person}{Alexander Ku}, {and}
  \bibinfo{person}{Dustin Tran}.} \bibinfo{year}{2018}\natexlab{}.
\newblock \showarticletitle{Image transformer}. In
  \bibinfo{booktitle}{\emph{International Conference on Machine Learning}}.
  PMLR, \bibinfo{pages}{4055--4064}.
\newblock


\bibitem[\protect\citeauthoryear{Pi, Bian, Zhou, Zhu, and Gai}{Pi
  et~al\mbox{.}}{2019}]%
        {pi2019practice}
\bibfield{author}{\bibinfo{person}{Qi Pi}, \bibinfo{person}{Weijie Bian},
  \bibinfo{person}{Guorui Zhou}, \bibinfo{person}{Xiaoqiang Zhu}, {and}
  \bibinfo{person}{Kun Gai}.} \bibinfo{year}{2019}\natexlab{}.
\newblock \showarticletitle{Practice on long sequential user behavior modeling
  for click-through rate prediction}. In \bibinfo{booktitle}{\emph{Proceedings
  of the 25th ACM SIGKDD International Conference on Knowledge Discovery \&
  Data Mining}}.
\newblock


\bibitem[\protect\citeauthoryear{Pi, Zhou, Zhang, Wang, Ren, Fan, Zhu, and
  Gai}{Pi et~al\mbox{.}}{2020}]%
        {pi2020search}
\bibfield{author}{\bibinfo{person}{Qi Pi}, \bibinfo{person}{Guorui Zhou},
  \bibinfo{person}{Yujing Zhang}, \bibinfo{person}{Zhe Wang},
  \bibinfo{person}{Lejian Ren}, \bibinfo{person}{Ying Fan},
  \bibinfo{person}{Xiaoqiang Zhu}, {and} \bibinfo{person}{Kun Gai}.}
  \bibinfo{year}{2020}\natexlab{}.
\newblock \showarticletitle{Search-based User Interest Modeling with Lifelong
  Sequential Behavior Data for Click-Through Rate Prediction}. In
  \bibinfo{booktitle}{\emph{Proceedings of the 29th ACM International
  Conference on Information \& Knowledge Management}}.
\newblock


\bibitem[\protect\citeauthoryear{Qin, Zhang, Wu, Jin, Fang, and Yu}{Qin
  et~al\mbox{.}}{2020}]%
        {qin2020user}
\bibfield{author}{\bibinfo{person}{Jiarui Qin}, \bibinfo{person}{Weinan Zhang},
  \bibinfo{person}{Xin Wu}, \bibinfo{person}{Jiarui Jin},
  \bibinfo{person}{Yuchen Fang}, {and} \bibinfo{person}{Yong Yu}.}
  \bibinfo{year}{2020}\natexlab{}.
\newblock \showarticletitle{User Behavior Retrieval for Click-Through Rate
  Prediction}. In \bibinfo{booktitle}{\emph{Proceedings of the 43rd
  International ACM SIGIR Conference on Research and Development in Information
  Retrieval}}.
\newblock


\bibitem[\protect\citeauthoryear{Ren, Qin, Fang, Zhang, Zheng, Bian, Zhou, Xu,
  Yu, Zhu, et~al\mbox{.}}{Ren et~al\mbox{.}}{2019}]%
        {ren2019lifelong}
\bibfield{author}{\bibinfo{person}{Kan Ren}, \bibinfo{person}{Jiarui Qin},
  \bibinfo{person}{Yuchen Fang}, \bibinfo{person}{Weinan Zhang},
  \bibinfo{person}{Lei Zheng}, \bibinfo{person}{Weijie Bian},
  \bibinfo{person}{Guorui Zhou}, \bibinfo{person}{Jian Xu},
  \bibinfo{person}{Yong Yu}, \bibinfo{person}{Xiaoqiang Zhu}, {et~al\mbox{.}}}
  \bibinfo{year}{2019}\natexlab{}.
\newblock \showarticletitle{Lifelong sequential modeling with personalized
  memorization for user response prediction}. In
  \bibinfo{booktitle}{\emph{Proceedings of the 42nd International ACM SIGIR
  Conference on Research and Development in Information Retrieval}}.
\newblock


\bibitem[\protect\citeauthoryear{Rendle, Freudenthaler, and
  Schmidt-Thieme}{Rendle et~al\mbox{.}}{2010}]%
        {rendle2010factorizing}
\bibfield{author}{\bibinfo{person}{Steffen Rendle}, \bibinfo{person}{Christoph
  Freudenthaler}, {and} \bibinfo{person}{Lars Schmidt-Thieme}.}
  \bibinfo{year}{2010}\natexlab{}.
\newblock \showarticletitle{Factorizing personalized markov chains for
  next-basket recommendation}. In \bibinfo{booktitle}{\emph{Proceedings of the
  19th international conference on World wide web}}. \bibinfo{pages}{811--820}.
\newblock


\bibitem[\protect\citeauthoryear{Sabour, Frosst, and Hinton}{Sabour
  et~al\mbox{.}}{2017}]%
        {sabour2017dynamic}
\bibfield{author}{\bibinfo{person}{Sara Sabour}, \bibinfo{person}{Nicholas
  Frosst}, {and} \bibinfo{person}{Geoffrey~E Hinton}.}
  \bibinfo{year}{2017}\natexlab{}.
\newblock \showarticletitle{Dynamic Routing Between Capsules}. In
  \bibinfo{booktitle}{\emph{Advances in Neural Information Processing
  Systems}}, Vol.~\bibinfo{volume}{30}.
\newblock


\bibitem[\protect\citeauthoryear{Sun, Liu, Wu, Pei, Lin, Ou, and Jiang}{Sun
  et~al\mbox{.}}{2019}]%
        {sun2019bert4rec}
\bibfield{author}{\bibinfo{person}{Fei Sun}, \bibinfo{person}{Jun Liu},
  \bibinfo{person}{Jian Wu}, \bibinfo{person}{Changhua Pei},
  \bibinfo{person}{Xiao Lin}, \bibinfo{person}{Wenwu Ou}, {and}
  \bibinfo{person}{Peng Jiang}.} \bibinfo{year}{2019}\natexlab{}.
\newblock \showarticletitle{BERT4Rec: Sequential recommendation with
  bidirectional encoder representations from transformer}. In
  \bibinfo{booktitle}{\emph{Proceedings of the 28th ACM international
  conference on information and knowledge management}}.
  \bibinfo{pages}{1441--1450}.
\newblock


\bibitem[\protect\citeauthoryear{Tang and Wang}{Tang and Wang}{2018}]%
        {tang2018personalized}
\bibfield{author}{\bibinfo{person}{Jiaxi Tang} {and} \bibinfo{person}{Ke
  Wang}.} \bibinfo{year}{2018}\natexlab{}.
\newblock \showarticletitle{Personalized top-n sequential recommendation via
  convolutional sequence embedding}. In \bibinfo{booktitle}{\emph{Proceedings
  of the Eleventh ACM International Conference on Web Search and Data Mining}}.
  \bibinfo{pages}{565--573}.
\newblock


\bibitem[\protect\citeauthoryear{Vaswani, Shazeer, Parmar, Uszkoreit, Jones,
  Gomez, Kaiser, and Polosukhin}{Vaswani et~al\mbox{.}}{2017}]%
        {vaswani2017attention}
\bibfield{author}{\bibinfo{person}{Ashish Vaswani}, \bibinfo{person}{Noam
  Shazeer}, \bibinfo{person}{Niki Parmar}, \bibinfo{person}{Jakob Uszkoreit},
  \bibinfo{person}{Llion Jones}, \bibinfo{person}{Aidan~N Gomez},
  \bibinfo{person}{{\L}ukasz Kaiser}, {and} \bibinfo{person}{Illia
  Polosukhin}.} \bibinfo{year}{2017}\natexlab{}.
\newblock \showarticletitle{Attention is all you need}. In
  \bibinfo{booktitle}{\emph{Proceedings of the 31st International Conference on
  Neural Information Processing Systems}}.
\newblock


\bibitem[\protect\citeauthoryear{Wang, Girshick, Gupta, and He}{Wang
  et~al\mbox{.}}{2018}]%
        {wang2018non}
\bibfield{author}{\bibinfo{person}{Xiaolong Wang}, \bibinfo{person}{Ross
  Girshick}, \bibinfo{person}{Abhinav Gupta}, {and} \bibinfo{person}{Kaiming
  He}.} \bibinfo{year}{2018}\natexlab{}.
\newblock \showarticletitle{Non-local Neural Networks}. In
  \bibinfo{booktitle}{\emph{2018 IEEE/CVF Conference on Computer Vision and
  Pattern Recognition}}.
\newblock


\bibitem[\protect\citeauthoryear{Xiao, Yang, Jiang, Wei, Hu, and Wang}{Xiao
  et~al\mbox{.}}{2020}]%
        {xiao2020deep}
\bibfield{author}{\bibinfo{person}{Zhibo Xiao}, \bibinfo{person}{Luwei Yang},
  \bibinfo{person}{Wen Jiang}, \bibinfo{person}{Yi Wei}, \bibinfo{person}{Yi
  Hu}, {and} \bibinfo{person}{Hao Wang}.} \bibinfo{year}{2020}\natexlab{}.
\newblock \showarticletitle{Deep Multi-Interest Network for Click-through Rate
  Prediction}. In \bibinfo{booktitle}{\emph{Proceedings of the 29th ACM
  International Conference on Information \& Knowledge Management}}.
  \bibinfo{pages}{2265--2268}.
\newblock


\bibitem[\protect\citeauthoryear{Xu, Li, Cui, Huang, Wei, and Zhou}{Xu
  et~al\mbox{.}}{2020}]%
        {xu2020layoutlm}
\bibfield{author}{\bibinfo{person}{Yiheng Xu}, \bibinfo{person}{Minghao Li},
  \bibinfo{person}{Lei Cui}, \bibinfo{person}{Shaohan Huang},
  \bibinfo{person}{Furu Wei}, {and} \bibinfo{person}{Ming Zhou}.}
  \bibinfo{year}{2020}\natexlab{}.
\newblock \showarticletitle{Layoutlm: Pre-training of text and layout for
  document image understanding}. In \bibinfo{booktitle}{\emph{Proceedings of
  the 26th ACM SIGKDD International Conference on Knowledge Discovery \& Data
  Mining}}. \bibinfo{pages}{1192--1200}.
\newblock


\bibitem[\protect\citeauthoryear{Yu, Liu, Wu, Wang, and Tan}{Yu
  et~al\mbox{.}}{2016}]%
        {yu2016dynamic}
\bibfield{author}{\bibinfo{person}{Feng Yu}, \bibinfo{person}{Qiang Liu},
  \bibinfo{person}{Shu Wu}, \bibinfo{person}{Liang Wang}, {and}
  \bibinfo{person}{Tieniu Tan}.} \bibinfo{year}{2016}\natexlab{}.
\newblock \showarticletitle{A dynamic recurrent model for next basket
  recommendation}. In \bibinfo{booktitle}{\emph{Proceedings of the 39th
  International ACM SIGIR conference on Research and Development in Information
  Retrieval}}. \bibinfo{pages}{729--732}.
\newblock


\bibitem[\protect\citeauthoryear{Zhang, Tay, Yao, and Sun}{Zhang
  et~al\mbox{.}}{2018}]%
        {zhang2018next}
\bibfield{author}{\bibinfo{person}{Shuai Zhang}, \bibinfo{person}{Yi Tay},
  \bibinfo{person}{Lina Yao}, {and} \bibinfo{person}{Aixin Sun}.}
  \bibinfo{year}{2018}\natexlab{}.
\newblock \showarticletitle{Next item recommendation with self-attention}.
\newblock \bibinfo{journal}{\emph{arXiv preprint arXiv:1808.06414}}
  (\bibinfo{year}{2018}).
\newblock


\bibitem[\protect\citeauthoryear{Zhou, Mou, Fan, Pi, Bian, Zhou, Zhu, and
  Gai}{Zhou et~al\mbox{.}}{2019}]%
        {zhou2019deep}
\bibfield{author}{\bibinfo{person}{Guorui Zhou}, \bibinfo{person}{Na Mou},
  \bibinfo{person}{Ying Fan}, \bibinfo{person}{Qi Pi}, \bibinfo{person}{Weijie
  Bian}, \bibinfo{person}{Chang Zhou}, \bibinfo{person}{Xiaoqiang Zhu}, {and}
  \bibinfo{person}{Kun Gai}.} \bibinfo{year}{2019}\natexlab{}.
\newblock \showarticletitle{Deep interest evolution network for click-through
  rate prediction}. In \bibinfo{booktitle}{\emph{Proceedings of the AAAI
  conference on artificial intelligence}}, Vol.~\bibinfo{volume}{33}.
  \bibinfo{pages}{5941--5948}.
\newblock


\bibitem[\protect\citeauthoryear{Zhou, Zhu, Song, Fan, Zhu, Ma, Yan, Jin, Li,
  and Gai}{Zhou et~al\mbox{.}}{2018}]%
        {zhou2018deep}
\bibfield{author}{\bibinfo{person}{Guorui Zhou}, \bibinfo{person}{Xiaoqiang
  Zhu}, \bibinfo{person}{Chenru Song}, \bibinfo{person}{Ying Fan},
  \bibinfo{person}{Han Zhu}, \bibinfo{person}{Xiao Ma},
  \bibinfo{person}{Yanghui Yan}, \bibinfo{person}{Junqi Jin},
  \bibinfo{person}{Han Li}, {and} \bibinfo{person}{Kun Gai}.}
  \bibinfo{year}{2018}\natexlab{}.
\newblock \showarticletitle{Deep interest network for click-through rate
  prediction}. In \bibinfo{booktitle}{\emph{Proceedings of the 24th ACM SIGKDD
  International Conference on Knowledge Discovery \& Data Mining}}.
  \bibinfo{pages}{1059--1068}.
\newblock


\end{thebibliography}

\clearpage
\appendix

\section{Appendix}
\balance
\subsection{Dataset Pre-processing}\label{apendix:data}
For MovieLens datasets we treat the presence of a movie rating as implicit feedback (i.e., we consider it as an action in the user's behavior history). For all datasets, we drop users and items with less than 5 related actions. For the Industrial dataset, 1\% users are randomly sampled from the ones who are active on February 1, 2021, and their click histories are collected. The full user-item interaction data contains \numprint{71086374} items, and \numprint{20059521441} actions from \numprint{42995680} users. The negative items used in evaluation for each user are randomly sampled from the set of items that the user has not interacted with using a uniform distribution. We plan to open this dataset to further nourish the community development in this field.

\subsection{Parameter Settings}
As mentioned above, we implement our method and all the baselines with PyTorch. We use an efficient implementation of the linear attention mechanism~\cite{choromanski2020rethinking} provided by~\cite{likhosherstov2020sub} for our LimaRec. All models are trained with an embedding dimension of 32, and a dropout rate of 0.1. We set the maximum number of training epochs to 500 on the ML-1M dataset, 200 on the ML-25M and XLong datasets, and 20 for Industrial. The number of self-attention blocks are set to 2 for SASRec, and we employ a single attention head, as it is pointed out in~\cite{kang2018self} that using multiple attention heads does not contribute to the recommendation performance. The regulation coefficient is set to 0.01 for HPMN and MIMN. We tune the number of interests used in ComiRec and MIND, as well as the number of hierarchical GRU layers in HPMN from 1 to 9. The number of memory slots of MIMN are set to 3.

\subsection{Experiment Environment}
We implement all models with PyTorch, and conduct experiments on a server with NVIDIA V100 GPUs. Regarding software versions, Python 3.6.9 and PyTorch 1.8.0 are used.

\subsection{Ablation Study of Multi-Interest Extraction Module}
\subsubsection{Settings}
In this section, we conduct an ablation analysis to further study the effectiveness of the multi-interest extraction module, and to compare the performance of incremental attention with vanilla self-attention. Hence we evaluate the recommendation performance of LimaRec without the multi-interest extraction layer, by only using the two incremental self-attention blocks. The results are presented in Table~\ref{tab:ablation}.

\begin{table}[b]
    \centering
    \caption{Performance of LimaRec without the Multi-Interest Extraction Module.}
    \label{tab:ablation}
    \begin{tabular}{lcccc}
        \toprule
        & HR@5 & NDCG@5 & HR@10 & NDCG@10 \\
        \midrule
        ML-1M & 0.6548 & 0.5003 & 0.7803 & 0.5410 \\
        ML-25M & 0.9172 & 0.7735 & 0.9686 & 0.7904  \\
        XLong & 0.6648 & 0.5058 & 0.7917 & 0.5470 \\
        Industrial & 0.7518 & 0.6147 & 0.8463 & 0.6454  \\
        \bottomrule
    \end{tabular}
\end{table}

\subsubsection{Findings}
We find that LimaRec performs notably worse without the multi-interest extraction module, which further indicates the necessity to model and capture the multi-facet interests in users' lifelong sequences. Compare the results with SASRec-1000 in Table~\ref{tab:overall_rec_performance}, we observe that the performance of LimaRec without multi-interest modeling (note that in this case, LimaRec only utilizes the two incremental self-attention blocks) is only slightly worse than SASRec using lifelong sequences. This proves that the incremental attention mechanism indeed share the same power of vanilla to capture long-term dependencies in lifelong sequences, identify relevant information and filter out noise.

\end{document}